\documentclass[12pt,preprint]{aastex}

\shorttitle{Magnetohydrostatic Prominences}
\shortauthors{Petrie, Blokland \& Keppens}

\begin{document}


\title{Magnetohydrostatic solar prominences in near-potential coronal magnetic fields}


\author{G.J.D. Petrie$^1$, J.W.S. Blokland$^2$ \& R. Keppens$^{2,3,4}$}
\affil{$^1$National Solar Observatory, 950 N. Cherry Avenue, Tucson, AZ 85719\\
$^2$FOM-Institute for Plasma Physics, P.O. Box 1207, 3430 BE Nieuwegein, the Netherlands\\
$^3$Centre for Plasma Astrophysics, K.U. Leuven, 3001 Heverlee, Belgium\\
$^4$Astronomical Institute, Utrecht University, PO Box 80000, NL-3508 TA Utrecht, The Netherlands}



\begin{abstract}
We present numerical magnetohydrostatic solutions describing the gravitationally stratified, bulk equilibrium of cool, dense prominence plasma embedded in a near-potential coronal field.  These solutions are calculated using the FINESSE magnetohydrodynamics equilibrium solver and describe the morphologies of magnetic field distributions in and around prominences and the cool prominence plasma that these fields support.  The equilibrium condition for this class of problem is usually different in distinct subdomains, separated by free boundaries, across which solutions are matched by suitable continuity or jump conditions describing force balance.  We employ our precise finite element elliptic solver to calculate solutions not accessible by previous analytical techniques with temperature or entropy prescribed as free functions of the magnetic flux function, including a range of values of the polytropic index, temperature variations mainly across magnetic field lines and photospheric field profiles sheared close to the polarity inversion line.  Out of the many examples computed here, perhaps the most noteworthy is one which reproduces precisely the
three-part structure often encountered in observations: a cool dense prominence within a cavity/flux rope embedded in a hot corona.  The stability properties of these new equilibria, which may be relevant to solar eruptions, can be determined in the form of a full resistive MHD spectrum using a companion hyperbolic stability solver.
\end{abstract}



\section{Introduction}

Solar prominences are cool, dense concentrations of plasma suspended by magnetic fields about 10,000~km clear of the base of the corona (Tandberg-Hanssen~1995).  The prominence is referred to as a filament when it is observed in absorption against the solar disk.  Observing, understanding and modeling prominences and the equilibrium and stability properties of their magnetic
fields is a major topic of solar research.  Eventually many quiescent prominences erupt abruptly and are expelled through the corona into interplanetary space embedded within a coronal mass ejection (CME).  CMEs are the most powerful drivers of the Sun-Earth connection.  Observations
(e.g. Munro et al. 1979, Webb \& Hundhausen 1987) and recent developments in
solar magnetohydrodynamics have shown that more than 70\% of CMEs originate from
eruptions of prominences caused by
failures of the confinement of highly twisted magnetic fields (Low~1996, 2001, Amari et al. 2003a, 2003b; Fan \& Gibson 2003, 2004; Fan \& Low 2003; Flyer et al. 2004, 2005, Zhang \& Low 2004, 2005).   Recently, using magneto-friction models, Mackay \& van Ballegooijen~(2006) considered the formation and ejection of flux ropes as the result of the build-up and concentration of axial flux by a simple reconnection process.

The magnetic field is central to the formation and persistence of quiescent prominences (Tandberg-Hanssen~1995).  It provides support against gravity for the electrically highly conducting prominence plasma and also acts as a thermal shield for the cool prominence against the million-degree coronal environment.  Observations have shown that long quiescent prominences form from the remnants of decaying active regions (Gaizauskas et al.~2001), and tend to lie in a cavity at the base of a well-formed coronal helmet streamer, running along the polarity inversion line of an extensive bipolar region on the photosphere (Tandberg-Hanssen~1995).  These prominences also tend to be suspended higher in the corona ($>30,000$~km above the photosphere) than those in active regions, which tend to be shorter in length and more contorted in shape by the more complex active region fields.  We adopt the hypothesis of Low \& Hundhausen~(1995) that quiescent and active-region prominences share the same basic magnetic structure and mechanics although they differ in magnetic intensity, structural complexity, structural length scale and evolutionary time scale.

The long life of a quiescent prominence as a macroscopic structure, from days to a week or more, suggests that static equilibrium is a reasonable first approximation to describe the prominence (Anzer 1989).  We confine our equilibrium study to 2D configurations for two reasons.  Firstly, prominences are significantly larger in one horizontal direction than the other, being about 200,000~km long and about 5,000~km wide, so that their bulk properties can be modeled in 2D.  They are invariably found parallel to and above photospheric polarity inversion lines.  Our 2D models describe flux ropes levitating in the solar atmosphere above the photosphere, which is taken as an infinite plane.  The topology of the ends of the flux rope, which in reality curve down to meet the photosphere in a full 3D configuration, is left out of the 2D description.  This description captures features of flux ropes the main part of whose lengths are suspended in the atmosphere above the photosphere.

The second reason is that in 3D the MHS force-balance equation becomes highly nonlinear, and a follow-up stability analysis becomes even more challenging as the effect of line-tying at the photospheric end
points must be taken into account as well.  Furthermore, the 3D MHS force-balance equation changes type from elliptic to mixed type, causing the character of the boundary value problem to change.  The 2D MHS (Grad-Shafranov) equation has a linear, elliptic differential operator so that its characteristic curves are all imaginary.  Thus, like fields $\phi$ described by, e.g., Laplace's equation $\nabla^2\phi =0$ or the linear wave equation $\nabla^2\phi +k^2\phi =0$, solutions $\psi$ of the Grad-Shafranov equation are continuous functions uniquely determined by specifying their value, normal derivative or linear combination of the two on the boundary.  The situation with the 3D MHS problem is very different (Parker 1979, 1994): the inclusion of three-dimensional variations causes the force-balance equation to change type from a simple elliptic equation with purely imaginary characteristics to an equation of mixed type with both imaginary and real characteristics.   Across the real characteristic curves field derivatives may not be defined.  Since these real characteristics coincide with field trajectories, tangential discontinuities are permitted there provided that the total pressure $|{\bf B}|^2/8\pi +p$ is balanced across the discontinuity.  In general, equilibrium solvers break down for mixed-type problems and it can been questioned whether the equilibrium and perturbation can be meaningfully distinguished in these mixed-type regimes, as has been questioned for hyperbolic regimes in mixed-type problems (Goedbloed 2003, 2004). 

The stability properties of these new equilibria, which may be relevant to solar eruptions, can be determined in the form of a full resistive MHD spectrum by solving the linearized MHD equations using the companion hyperbolic stability solver PHOENIX (Blokland et al.~2007a).  MHD waves and instabilities control the dynamics of plasma and occur as the natural response to global excitation.  Measurement of the spectrum of MHD waves gives direct information on the internal state of the plasma.  This is called MHD spectroscopy in analogy with quantum mechanical spectroscopy, which also involves eigenvalue problems of linear operators.  MHD spectroscopy entails a separate study of the nonlinear static equilibrium configuration on the one hand and the various linear wave structures that can occur on the other.  The first study is the subject of this paper and the spectroscopy will be treated in a sequel.

The paper is organized as follows.  We discuss the morphology of prominences and their associated magnetic fields in Section~\ref{topology}.  The basic magnetohydrostatic equilibrium problem is introduced in Section~\ref{mhsprob} before the new models are presented in Section~\ref{models}.  We conclude with a discussion in Section~\ref{discussion}.

\section{Prominence magnetic field morphology}
\label{topology}

After a lapse of 20 years or so, there has been a renewed interest in the measurement of prominence (and coronal) magnetic fields (Judge~1998, Lin et al.~1998, 2000, Lopez Ariste \& Casini~2002, 2003, Trujillo Bueno~2003).  New techniques for observing magnetic fields of prominences on the solar disk (Lin et al.~1998, Trujillo Bueno~2003) allow many new possibilities for diagnosing prominences, including low-lying ones inaccessible to older methods which relied on the prominences being clearly visible above the limb.

Spectro-polarimetry has shown that a prominence is threaded by a nearly horizontal magnetic field whose principal component is along the long axis of the prominence.  Prominences are classified by whether their magnetic fields thread in the same (called normal topology) or opposite (inverse topology) compared to the underlying photospheric bipolar field.  These two classes imply distinct topologies for the coronal magnetic fields around prominences.  These two basic topologies, normal and inverse, were first captured in the models of Kippenhahn \& Schl\'uter~(1957) and Kuperus \& Raadu~1974), respectively.  The observed principal component of prominence magnetic fields along the long axes of prominences implies that fields around prominences are characterized by highly sheared or twisted magnetic topologies.

Statistical studies of prominence topologies suggest that inverse prominences are more common than normal prominences by a factor of about three, while inverse prominences sit higher in the atmosphere than normal prominences (Leroy et al.~1984, 1989, Bommier et al.~1994).  Most prominences associated with helmet streamers have the inverse configuration (Leroy et al.~1984).  In the 2D inverse topology, the prominence must sit in a two-flux magnetic system, one flux connecting the bipolar magnetic sources in the photosphere below and the other forming a rope which embeds the prominence and runs above and parallel to the photospheric polarity inversion line.  In 3D, single-flux models can produce inverse-topologies and helmet streamer structures (e.g. Aulanier \& D\'emoulin~1998, Mackay \& van Ballegooijen~2005, 2006).  The prominence flux rope is seen as a cavity in coronal white light observations and as a filament channel in the chromosphere.  For the inverse topology with the prominence magnetic field in the opposite direction to that of the bipolar photospheric field region below, the magnetic flux rope containing the prominence supports a part of the prominence weight by current attraction from above (Low \& Hundhausen~ 1995).  Low \& Zhang~(2002) suggested that the two loosely defined classes of impulsively and gradually accelerated CMEs, fast and slow CMEs, may be explained by the different interactions between expulsion dynamics and magnetic reconnections taking place in the different inverse and normal field topologies.

A solar prominence cavity or flux rope may be idealized as a 2D structure running above and parallel to the photospheric polarity inversion line in the ignored direction, the $x$-direction.  In the $y$-$z$ plane, it is represented by a finite region of closed contours of the magnetic flux function with cool plasma localized by gravitational stratification at the bottom of the rope.  The characteristic hydrostatic scale length of the cool prominence plasma is a few hundred km while a typical flux rope size is estimated to be 10~Mm or more.  While details of magnetic field structure in and around prominences have not been precisely determined by observations, Low \& Hundhausen~(1995) give a plausible sketch of how this field may be structured in their Figure~1.  The flux contours are compressed at the bottom by the cool, dense prominence plasma.

\begin{figure*}[h]
\begin{center}
\resizebox{0.49\hsize}{!}{\includegraphics*{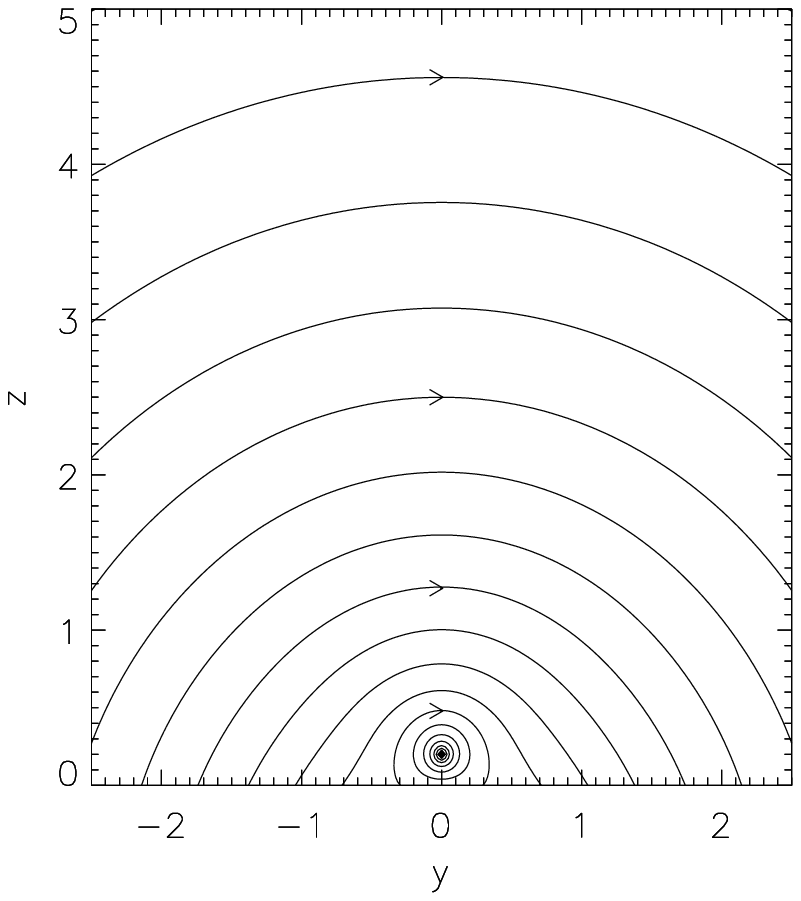}}
\resizebox{0.49\hsize}{!}{\includegraphics*{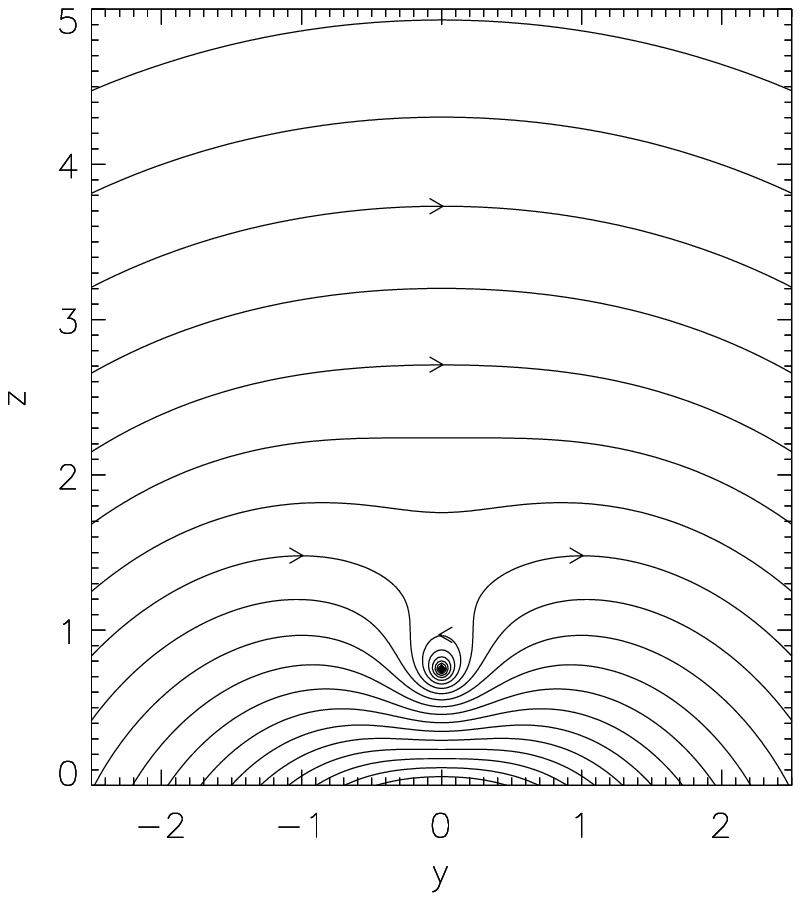}}
\end{center}
\caption{The inverse (left) and normal (right) prominence topologies for prominences represented by massive line currents.  In the inverse topology the line current flows in the same direction as the intrinsic photospheric current, while the currents are oppositely directed in the normal topology.  The Lorentz force is directed upward in both examples, as it must be to balance the weight.}
\label{topologies}
\end{figure*}

Amari \& Aly~(1989, 1992) modeled prominences by embedding massive MHS line currents and current sheets in twisted 2D force-free magnetic fields for the first time.  Then Low \& Hundhausen~(1995) gave an insightful model of the two topologies of quiescent prominences using line currents.  Taking the plane $z=0$ to be the photosphere, the magnetic field above this plane around the prominence mass, which is formed along a line $(y,z)=(0,z_2)$ parallel to the $x$-axis, with $z_2>0$, is given by the magnetic flux function

\begin{equation}
\psi =\frac{I_{\rm photo}}{c}\log [y^2+(z-z_1)^2]+\frac{I_{\rm prom}}{c}\log\left[\frac{y^2+(z-z_2)^2}{y^2+(z+z_2)^2}\right] ,\label{linecurrents}
\end{equation}

\noindent where $c$ is the speed of light.  The $I_{\rm photo}$ term causes the photospheric field to be bipolar and is due to a subphotospheric virtual line current at $(y,z)=(0,z_1)$ with $z_1<0$.  The $I_{\rm prom}$ term represents the line current associated with the prominence itself as well as an image current beneath the photosphere which causes this part of the flux function to vanish at the photosphere.  Thus, the vertical magnetic field $B_z$ at the photosphere is entirely unaffected by the presence of the filament-image pair, which only changes the horizontal magnetic field $B_y$.  There is no axial magnetic field component $B_x$ in this simple model, which is useful for investigating the magnetic field topology around the prominence and does not resolve the prominence structure itself.  For this simple line-current model, the two basic magnetic prominence topologies depend on whether $I_{\rm photo}$ and $I_{\rm prom}$ are of the same sign (inverse topology) or opposite signs (normal topology).  Figure~\ref{topologies} shows an example of each topology.  We will use this simple model to guide our numerical modeling in later sections.

Hundhausen \& Low~(1994) presented an analytical continuation method for calculating polytropic MHS equilibrium states in the Cartesian plane.  This method was exploited in Low \& Hundhausen~(1995)'s study of inverse prominences.  Whereas Low \& Hundhausen~(1995) represented prominence plasma enhancements by massive line currents and current sheets, Low \& Zhang~(2004) gave analytical solutions describing plasma enhancements of finite width by addressing a free-boundary problem.  By this approach they were able to describe both normal and inverse magnetic topologies.  This was made possible by the assumption of a circular boundary between the prominence plasma and the ambient corona and by making assumptions simplifying the polytropic equation of state.

In this paper we present classes of prominence-like solutions calculated numerically using the FINite Element Solver for Stationary Equilibria (FINESSE, Belien et al.~2002) which allows the computation of stationary, gravitationally stratified axisymmetric or cartesian configurations.  The two codes FINESSE and PHOENIX (Blokland et al.~2007a) have already been used in tandem to determine the stability of tokamaks and accretion disks with steady flows (Goedbloed et al.~2004a, b, Blokland et al.~2007b) but have not been adapted for the study of solar phenomena although the PARIS code (Belien et al.~1997) which treats the gravitational stratification of axisymmetric static loops is incorporated in the functionality of FINESSE.  Here, we use a suitably improved version of the FINESSE code to extend recent analytic work on translationally invariant magnetohydrostatic equilibria with uniform gravity in Cartesian geometry, by allowing a free boundary between the prominence flux rope and the ambient coronal magnetic field and allowing one to choose the polytropic index freely.  The various classes of magnetohydrostatic solutions discussed in what follows each lead to a second order
PDE, which FINESSE solves in weak form using a Picard iteration. We implemented the various forms
obtained for this PDE under different choices of freely chosen flux functions, along with their scalings,
as discussed in the Appendix.  We restrict our numerical calculations to static solutions which are translationally symmetric, all of which fall into an elliptic regime where a split between the equilibrium and the perturbations in a forthcoming stability analysis can meaningfully be performed.

\section{The magnetohydrostatic problem}
\label{mhsprob}

Consider the static-equilibrium model based on the
one-fluid ideal hydromagnetic description, denoting the magnetic field,
plasma pressure and density by ${\bf B}$, $p$ and $\rho$, respectively.
The balance of forces is described by

\begin{equation}
\label{fb}
{1 \over 4 \pi} \left(\nabla \times {\bf B}\right) \times {\bf B}
- \nabla p - \rho g {\bf\hat z} = 0 ,
\end{equation}

\noindent
assuming a uniform local gravity of acceleration $g$ in the $-z$ 
Cartesian direction.  Then the ideal gas law relates the gas pressure $p$ to the gas density $\rho$

\begin{equation}
\label{ideal}
p = \frac{k_B}{\mu} \rho T ,
\end{equation}

\noindent
where $k_B$ is Boltzmann's constant and $\mu$ is the mean particle mass for a fully ionized (monatomic) 
hydrogen plasma.  The solenoidal condition

\begin{equation}
\label{solenoid}
\nabla \cdot{\bf B} = 0 , 
\end{equation}

\noindent
closes the set of equations to determine $p$, $\rho$, and ${\bf B}$.  To keep the physical problem simple, we 
avoid the complication of a full energy equation by applying in turn two assumptions: (1)
that the plasma temperature is a flux function $T=T(\psi )$ (including the isothermal case $T=T_0$ a constant), and (2) the polytropic case where the entropy $s=p/\rho^{\gamma}=s(\psi )$ is a flux function.  (The polytropic case with $\gamma$ equal to the ratio of specific heats describes an adiabatic gas while the isothermal case eliminates heat transport.)  The value of the polytropic index, denoted here by $\gamma$, is not
well known in the corona. Parker~(1962)
gives $\gamma\approx 1.1-1.2$ as a guide and this seems to hold 40 years
after publication. Any value of $\gamma$ less than 5/3, the adiabatic lapse value, is
of potential physical interest.  The $\gamma < 1$ cases extend the work of Low \& Hundhausen~(1995) and Low \& Zhang~(2004)
in which $\gamma$ is of the form $\gamma = n/(n+1)$ where $n$ is
an integer which is positive in their cases.

In the 2D case with axial symmetry (${\partial /\partial x}=0$), the magnetic field satisfying equation~(\ref{solenoid}) has the form

\begin{equation}
{\bf B}=\left( f(y,z), \frac{\partial\psi (y,z)}{\partial z} , -\frac{\partial\psi (y,z)}{\partial y}\right) ,
\label{Bfield}
\end{equation}

\noindent
 where $f(y,z)$ is the axial magnetic field component and $\psi (y,z)$ is the magnetic flux function, whose isosurfaces contain the magnetic field trajectories.  By standard theory (Low 1975), the momentum equation~(\ref{fb}) reduces to

\begin{eqnarray}
\nabla^2\psi +f(\psi )\frac{df(\psi )}{d\psi} +4\pi\frac{\partial p(\psi , z)}{\partial\psi} & = & 0,\label{reducedfb}\\
\left.{\frac{\partial p(\psi, z)}{\partial z}}\right|_{\psi =\rm{const}} +\rho g & = & 0\label{hydrostatic} .
\end{eqnarray}

\noindent
Here $f(y,z)=f(\psi )$ must be a strict function of $\psi$ from the projection of the momentum equation on the symmetry axis.  Here we take the partial $z$ derivatives while keeping $\psi$ constant.
Equation~(\ref{hydrostatic}) is solved by

\begin{equation}
p(\psi ,z)=p_1(\psi )-\int_0^z \rho (\psi ,z')g dz'.
\label{pintegral}
\end{equation}

Details of the method used to solve Equation~(\ref{reducedfb}) numerically are given in the Appendix.  For simplicity a Cartesian grid with exponentially
varying gridpoint spacing is used to capture the detailed structure near the
symmetry axis. Thus gridpoints are packed
around the origin while the domain can extend to 50 units
or more, at which distances the field is rather insensitive to the details of the structure near the origin.

The solutions are calculated numerically, solving in both
potential and non-potential regions simultaneously,
exploiting the freedom to define $p_0 (\psi )$ and $f^2(\psi )$ arbitrarily.
  Piecewise definitions of $p_0$ and $f^2$ allow one to
calculate a full magnetohydrostatic equilibrium embedded in a potential
field at once.

The mapping from the Cartesian space $(y,z)$ onto the $(\psi ,z)$ space is not one-one in general because there may be more than one field line of the same value $\psi =\psi_c $ at a given height $z$.  The field lines in the inverse configuration typically have two points at a given height, or one in the half-domain $y>0$ in which the problem is solved in practice.  Then under symmetry about the $z$-axis, single functions $p(\psi ,z)$ and $\rho (\psi ,z)$ characterize the gas pressure and density fully.  Field lines of the normal topology, on the other hand, may have four (two in $y>0$) distinct points at the same height.  Moreover, the same constant value of $\psi$ may be found at two field lines: one anchored to the base of the atmosphere and one within the levitating flux rope.  Even with symmetry about the $z$-axis, the two points at the same height are generally complicated by having distinct gas pressures while they are mapped onto the same point $(\psi ,z)$.  In such a situation, $p(\psi ,z)$ and $\rho (\psi ,z)$ are multivalued functions of $\psi$ (Low \& Zhang~2004).

In general, the Grad-Shafranov equation can assume as many different forms as there are distinct regions of physical space, each occupied by a different branch of the functions $p(\psi ,z)$ and $\rho (\psi ,z)$.  The distinct regions are separated by free boundaries, the loci of which are unknown and must be calculated simultaneously with the solution $\psi$.  The free boundaries are governed by continuity or jump conditions imposed by force balance at these boundaries.  Free boundaries also separate regions where the functional forms of $p(\psi ,z)$ and $\rho (\psi ,z)$ differ, e.g., a region of potential or force-free magnetic field separated from a region of full MHS force balance.
 
 \section{The models}
 \label{models}
\subsection{Nonlinear force-free fields}
\label{nlff}

Before describing models with MHS force balance we discuss force-free flux-rope solutions.  The force-free case is the one with $p_0 (\psi ) =$~constant so that all Maxwell stresses are contained within the magnetic field and all electric currents are aligned with the magnetic field.  The simplest case, the potential case $f^2(\psi )=0$ cannot have closed magnetic flux contours by the maximum principle for the Laplacian differential operator.  Electric currents are necessary for a flux rope to be present.  The current associated with the magnetic field of Equation~(\ref{Bfield}) is described by

\begin{equation}
{\bf j}=\left( -\nabla^2\psi , \frac{d f(\psi)}{d\psi}\frac{\partial\psi}{\partial z} , -\frac{d f(\psi)}{d\psi}\frac{\partial\psi}{\partial y} \right) ,
\end{equation}

\noindent
and the linear case $f(\psi )=\alpha_0\psi$, for some constant $\alpha_0$, is seen to yield the usual Helmholtz equation for linear force-free fields $(\nabla^2 +\alpha_0^2)\psi =0$ - see also Equation~(\ref{reducedfb}).  The general condition for the nonlinear field to be force-free is that Equation~(\ref{reducedfb}) is satisfied with $\partial p(y,z)/\partial z = 0$, in which case ${\bf j}=\alpha (y,z){\bf B}$ with the nonlinear force-free parameter $\alpha (y,z)=df(\psi )/d\psi $.  The field-aligned currents are clearly associated with axial magnetic flux $B_x$ which is responsible for twist and shear in the magnetic field in these 2D solutions.

\begin{figure*}[ht]
\begin{center}
\resizebox{0.75\hsize}{!}{\includegraphics*{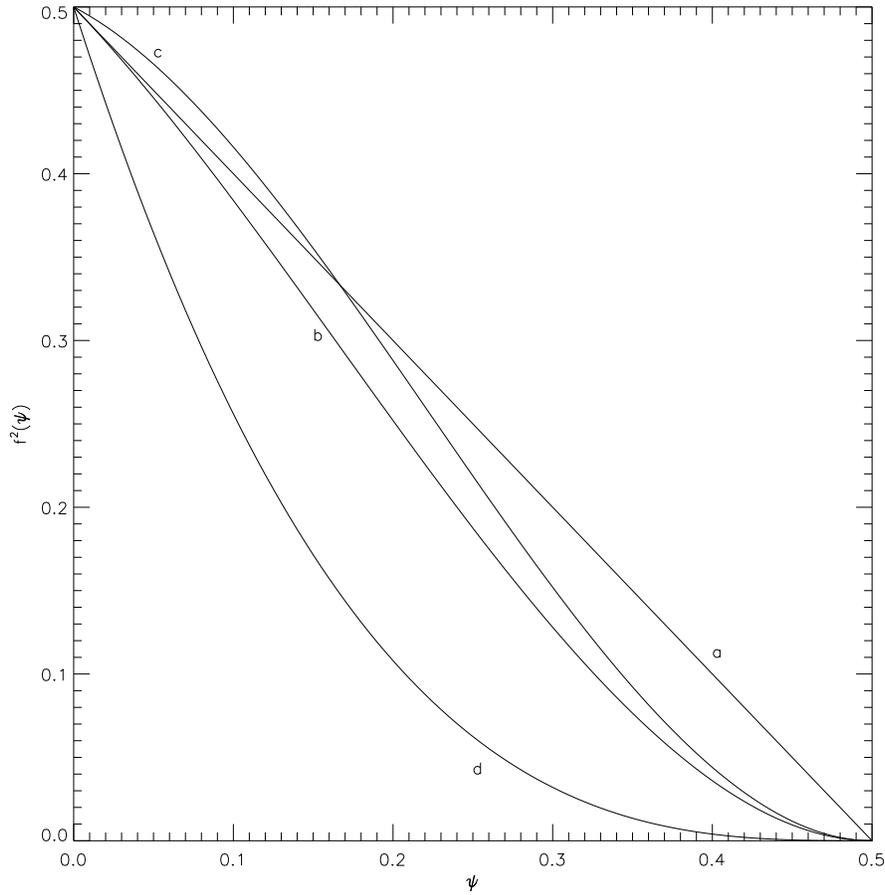}}
\end{center}
\caption{Examples of $f^2 (\psi )$ functions applied in this paper.  The quantity $f(\psi )$ is the axial component of $\bf B$, $B_x$.  For $\psi >0.5$ the profiles are zero.}
\label{Pi1s}
\end{figure*}

\begin{figure*}[ht]
\begin{center}
\resizebox{0.37\hsize}{!}{\includegraphics*{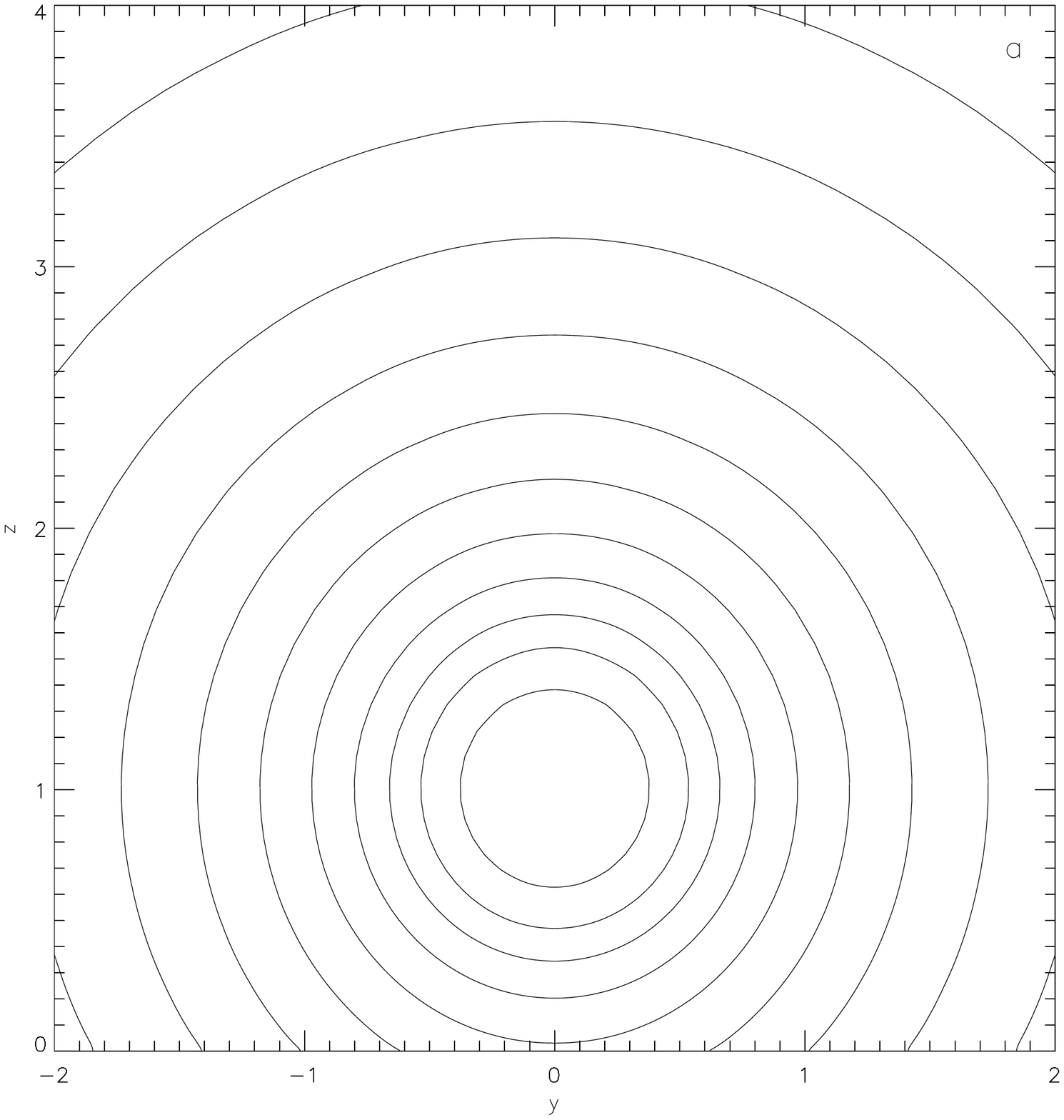}}
\resizebox{0.37\hsize}{!}{\includegraphics*{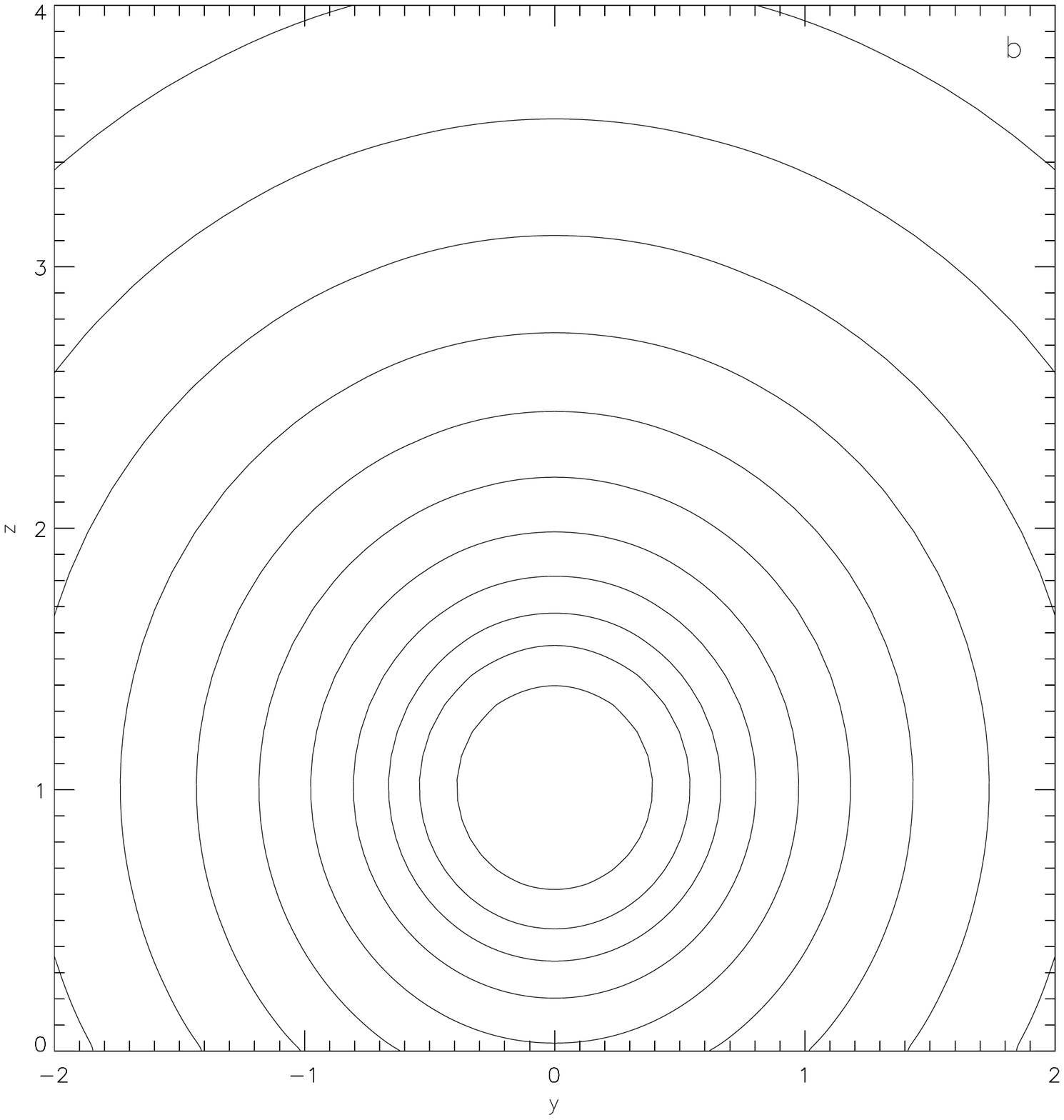}}
\resizebox{0.37\hsize}{!}{\includegraphics*{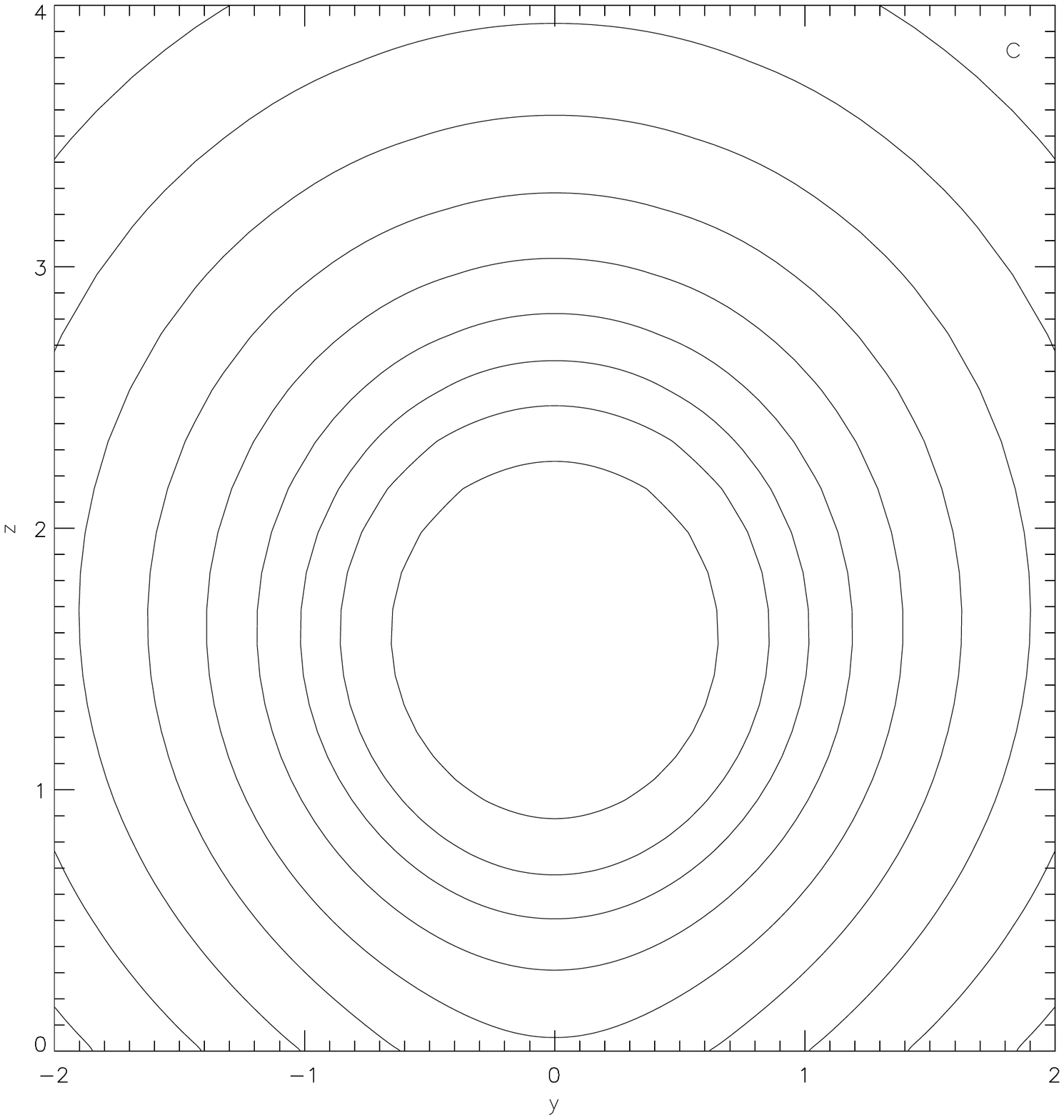}}
\resizebox{0.37\hsize}{!}{\includegraphics*{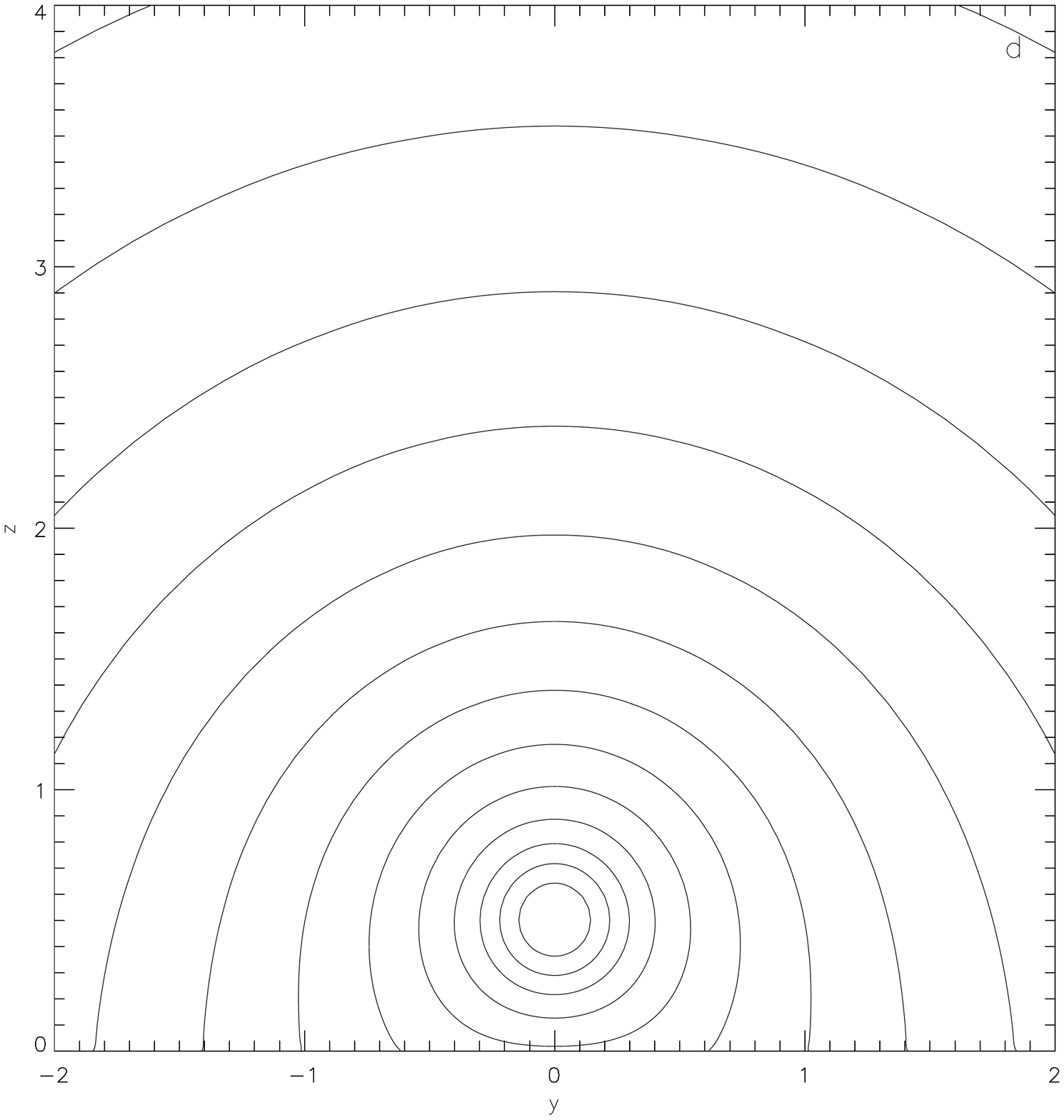}}
\end{center}
\caption{Force-free equilibria of inverse-polarity prominence magnetic fields with $f^2$ given by the curve in Figure~\ref{Pi1s} with label a (top left), b (top right), c (bottom left) and d (bottom right).}
\label{forcefree}
\end{figure*}

\begin{figure*}[ht]
\begin{center}
\resizebox{0.37\hsize}{!}{\includegraphics*{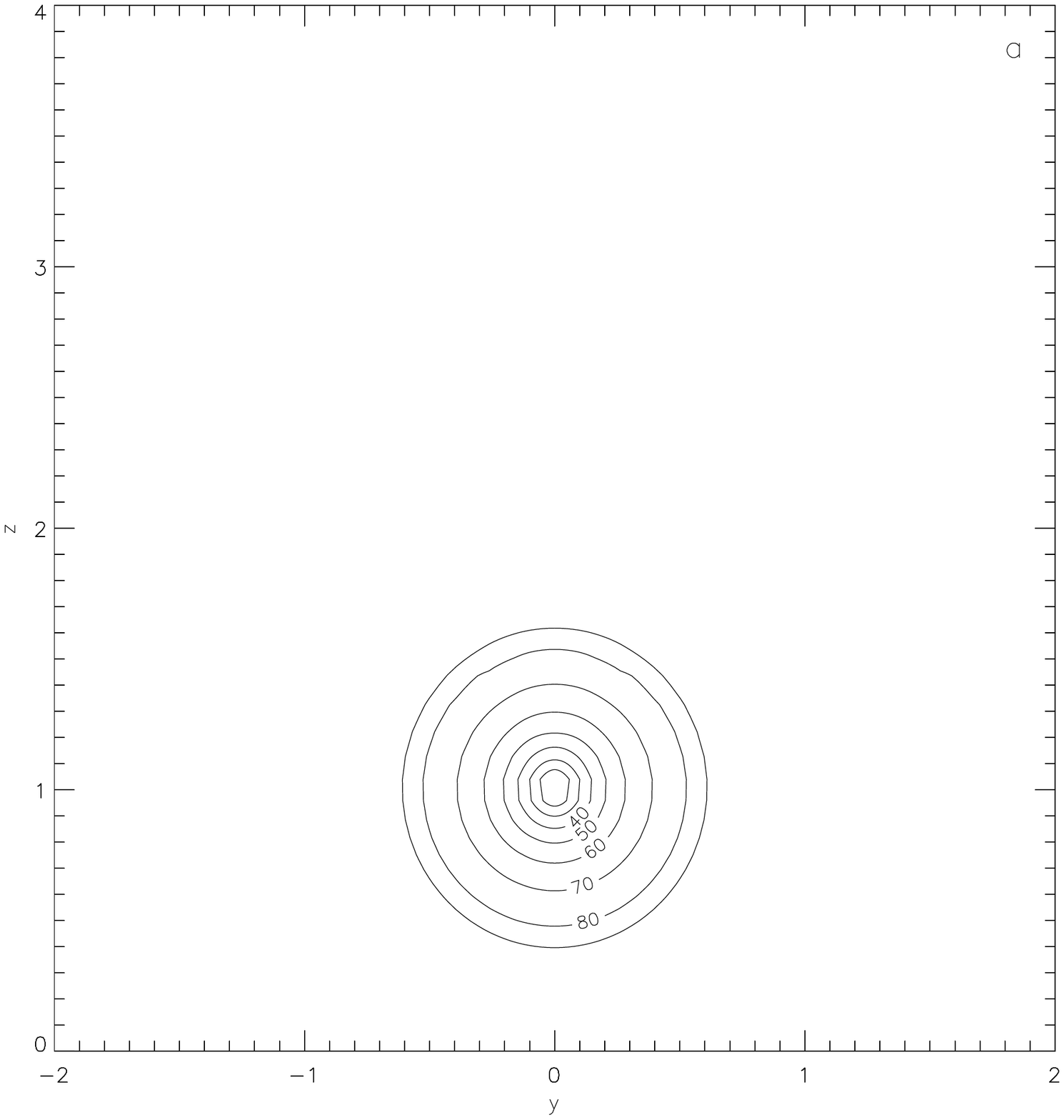}}
\resizebox{0.37\hsize}{!}{\includegraphics*{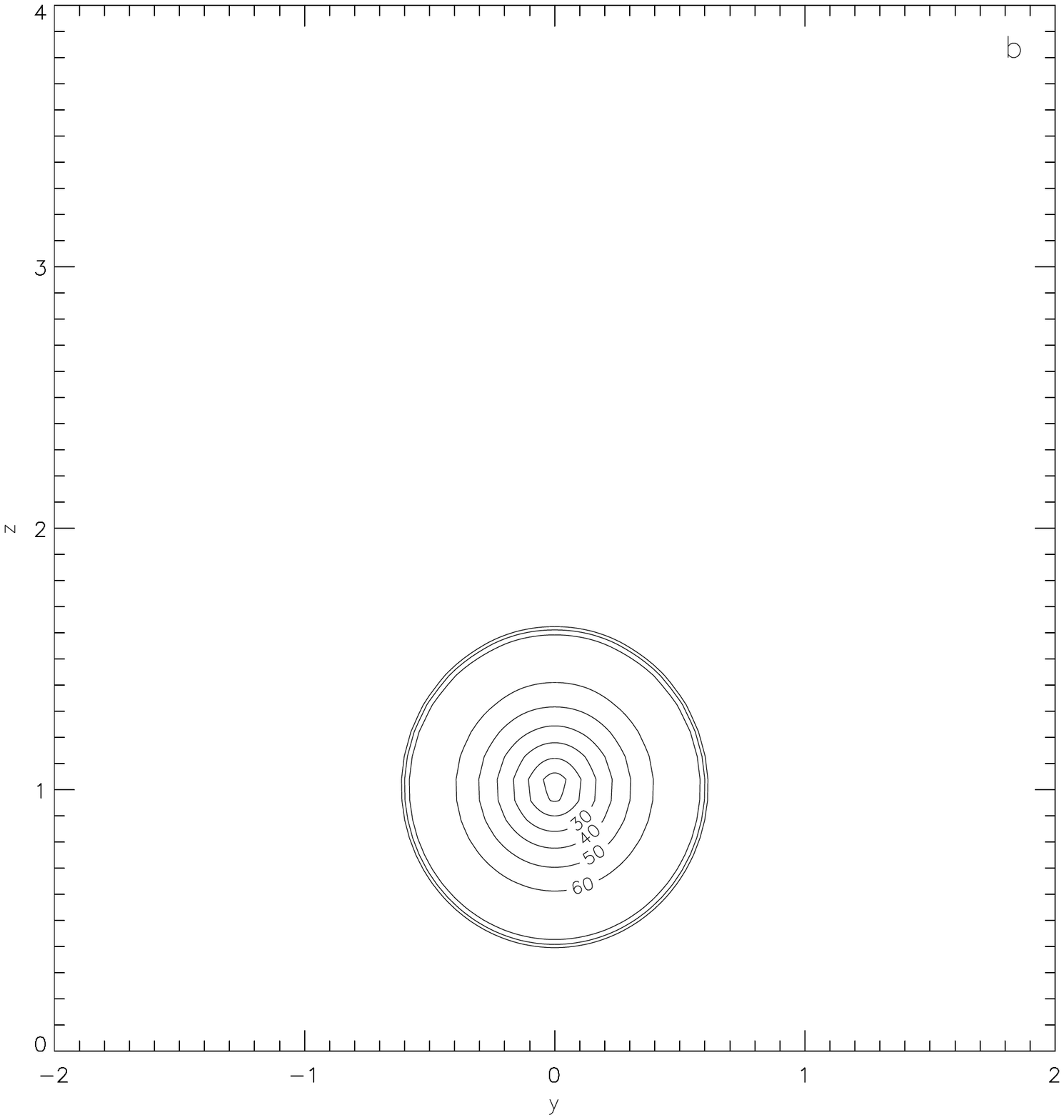}}
\resizebox{0.37\hsize}{!}{\includegraphics*{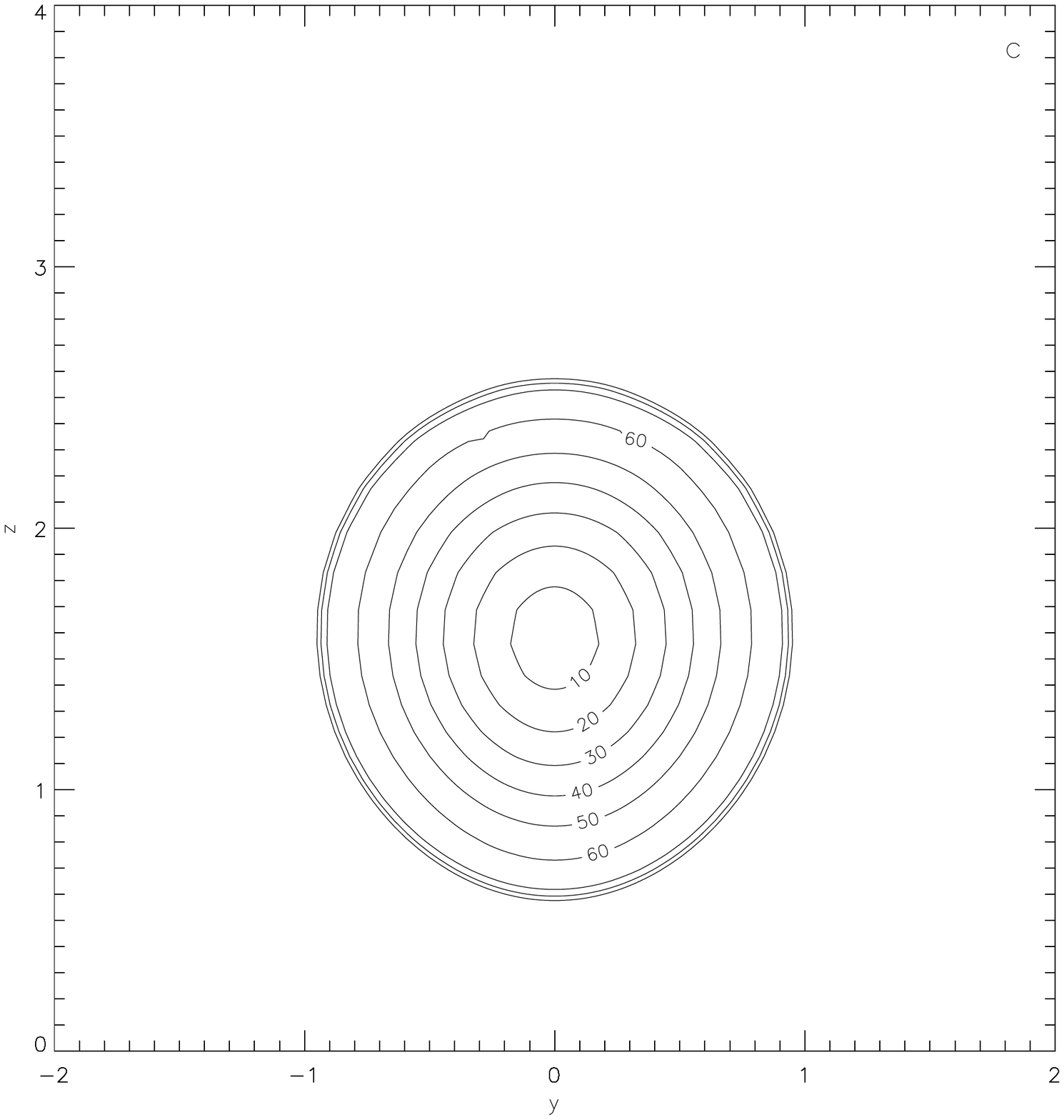}}
\resizebox{0.37\hsize}{!}{\includegraphics*{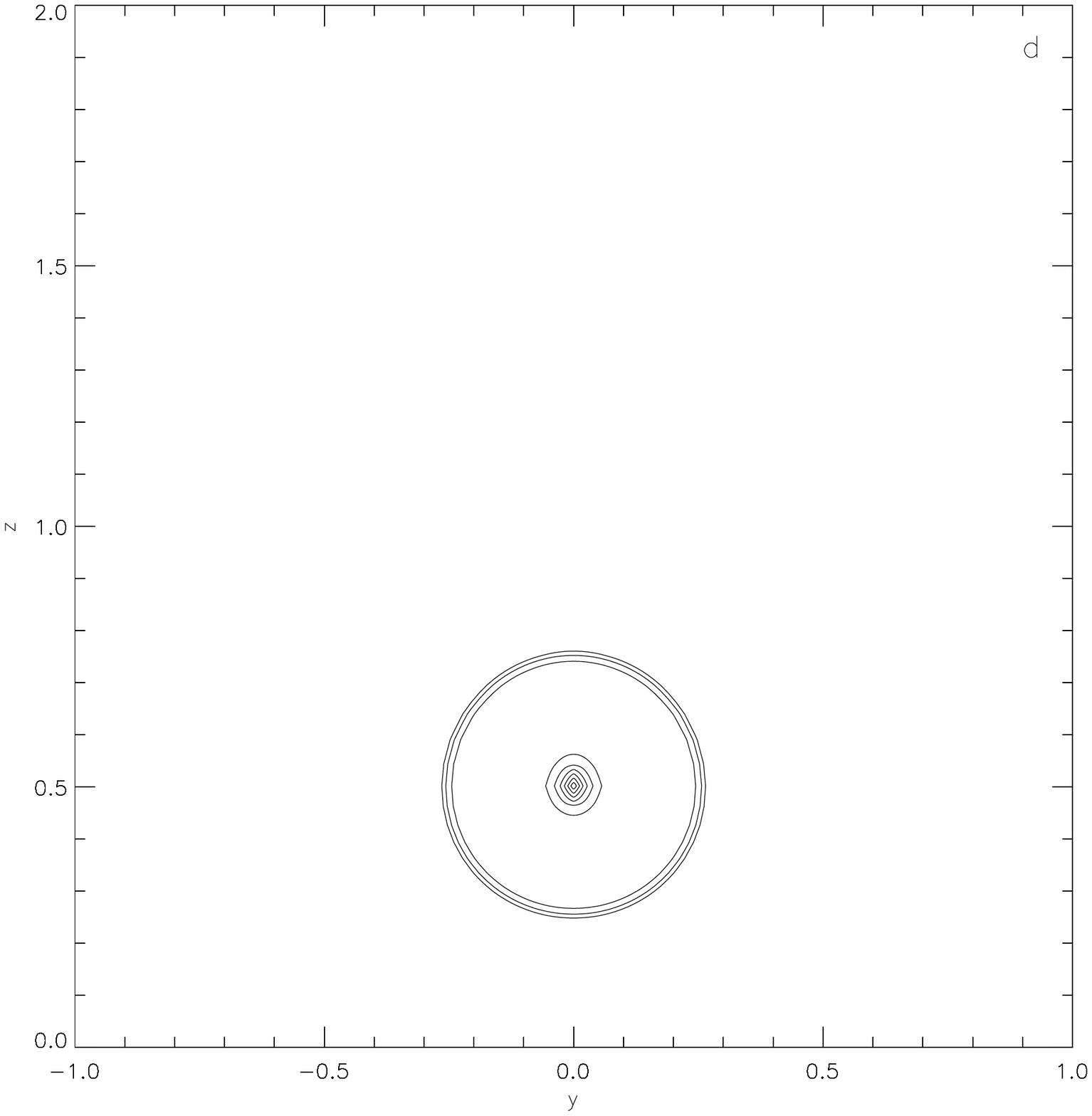}}
\end{center}
\caption{Force-free equilibria of inverse-polarity prominence magnetic fields: angles between the field vectors and magnetic axes in the four models of Figure~\ref{forcefree}.  The labels a, b, c and d match those in Figures \ref{Pi1s} and \ref{forcefree}.}
\label{angles}
\end{figure*}

If we insist that the ambient magnetic field be near-potential then the current distribution must be concentrated in the vicinity of the flux rope.  Therefore, in our examples $f^2(\psi )$ is non-zero only in regions bounded by a chosen closed flux contour (here $\psi =0.5$).  The observed increase of magnetic field strength and emission intensity to maxima at the
top of a prominence (Rust 1967, House \& Smart 1982,
Leroy et al.~1989) suggest forms of $f(\psi )$ and $p_0 (\psi )$ with maxima at the flux rope center.  In Low \& Hundhausen~(1995) and Low \& Zhang~(2004), $f^2$ is of the form $1/2 -\psi$, taking $\psi =0$ at the magnetic axis at the center of the flux rope and $\psi$ strictly positive elsewhere.  This profile is shown in Figure~\ref{Pi1s} as curve a.  Outside the contour $\psi =0.5$, $f^2 (\psi )\equiv 0$.   Figure~\ref{Pi1s} also shows several profiles for $f^2 (\psi )$.  Curves b, c and d are all of the form

\begin{equation}
f^2 (\psi )=\frac{1}{2} (1-2\psi)^2(1+n\psi ) ,
\label{Pi1eq}
\end{equation}

\noindent
with $n=2$, $3$ and $-2$ respectively.  There functions are the Hermite polynomials with double roots at $\psi =0.5$ and satisfying the conditions at $\psi =0$ $f^2 =0.5$ and $d f^2 /d\psi = -1$, $-1/2$ and $-3$ respectively.  The double root at $\psi =0.5$ allows a smooth transition from the current-carrying flux rope to the current-free ambient field.  The corresponding flux contours for the four cases are shown in Figure~\ref{forcefree} and the angle of the magnetic field with respect to the magnetic axis is shown for each case as functions of space in Figure~\ref{angles}.  The angle must be zero at $\psi =0$ since $B_y$ and $B_z$ must vanish there.  Outside the $\psi =0.5$ contour, the angle must be $90^{\circ}$ since $B_x=0$ there.  Therefore in each case the angle must vary from $0^{\circ}$ to $90^{\circ}$ within the flux rope.  The value of angle that an observer would find from one of our models depends on the nature of this variation from $0^{\circ}$ to $90^{\circ}$ and on the distribution of plasma within the flux rope.  Typical measured angles are of order $30^{\circ}$ (Leroy et al.~1984, 1989, Bommier et al.~1994).

Magnetic field models corresponding to the four $f^2$ profiles of Figure~\ref{Pi1s} are shown in Figure~\ref{forcefree}.  The boundary conditions are given by Equation~(\ref{linecurrents}) describing the solutions shown in Figure~\ref{topologies}.  These boundary conditions are motivated by the fact that a
(2D) flux rope resembles from a distance a line current and because of the relationship of this work to Low \&
Hundhausen~(1995), whose models the present work extends.  The $I_{\rm prom}$ term has no contribution to the photospheric boundary condition while its influence on the flux rope solutions when applied at the distant lateral and top boundaries is found to be effectively negligible except in extreme cases.  The solutions in this paper are therefore determined by the $I_{\rm photo}$ boundary condition, a simple bipole, and the form of the governing equation.  The subphotospheric line current modeled by $I_{\rm photo}$ provides a convenient means of producing the bipolar boundary conditions.  Alternative bipolar boundary conditions would produce similar results.

The four plots in Figure~\ref{forcefree} are labeled a, b, c and d to match the four curves in Figure~\ref{Pi1s}.  Curve c has a shallower decrease and curve d a steeper decrease in axial flux as one moves away from the magnetic axis than curve a, while curve b has the same decrease.  Models a and b are very similar indeed.  There are differences but they are too small to see in the plots.  Model c is larger than a and b while model d is smaller.  These effects are due to the different quantities of axial magnetic flux confined close to the magnetic axes of the models.  Figure~\ref{angles} shows the angle between the magnetic vector and the magnetic axis in each model.  Here, the difference between models a and b is much more obvious.  The magnetic angle of model b has a significantly sharper change at the edge of the flux rope than is the case for model a, and its angle is consistently smaller over much of the flux rope volume, indicating a more sheared field.  In model d almost all of the angle changes are concentrated at the center and the edge of the flux rope.  Here there is a concentration of highly sheared field at the center of the flux rope, outside which the angle has a plateau between $60^{\circ}$ and $70^{\circ}$, finally increasing sharply to $90^{\circ}$ at the edge.  In the other models the distribution of angle changes is more even.  The clumping of angle contours near the edge of each flux rope in models b, c and d is an effect of the total field strength there - small changes in $f^2$ cause large changes in the angle there.

Chromospheric fibrils as observed in H$\alpha$ are believed to delineate magnetic flux tube trajectories emanating from the photosphere.  Over the neighborhood of an active region these fibrils follow an organized pattern: approaching the filament from either side, the fibrils become longer and aligned more parallel to the filament axis.  Far from the filament the fibrils are nearly perpendicular to the filament, while at either side of the filament they are parallel.  No fibrils close to the polarity inversion line are observed to arc over it, and the fibrils indicate that magnetic flux in the positive and negative magnetic regions is directed away from and towards the inversion line respectively.  This is all contrary to what is expected of a near-potential magnetic field and indicates the presence of significant electric currents in the vicinity of as well as within the filaments.  Low \& Hundhausen~(1995) and Low \& Zhang~(2004) included photospheric magnetic shear in their models via their current-carrying fluxrope field.  If this field is allowed to dip beneath the photosphere, the solution has a sheared photospheric signature similar to the sheared photospheric polarity inversion lines associated with prominences in observations.  The same effect can be produced by our models by setting the photosphere to be just above $z=0.5$.

We also tried to find examples where the direction of the axial flux reverses.  One immediate motivation for doing this is the observation by
Lin et al.~(1998)  of a prominence in absorption against the disk
in mid-latitude with axial field component of opposite signs on opposite
sides of the photospheric neutral line below the prominence.  Flux rope
configurations resembling the reversed-field pinch (Freidberg 1987,
Biskamp 1993) familiar to plasma laboratory physicists would exhibit
precisely this pattern.  In Taylor's (1986) theory of plasma relaxation
to minimum energy states under conservation of magnetic helicity, there
exists a helicity threshold above which the relaxed field must exhibit a
reversal of its axial field across the tube cross-section, a result
repeatedly confirmed by experiment.  In
the more complex, open-state environment of the solar atmosphere it is not clear if corresponding phenomena are to be expected.  We were not able to obtain to obtain cases with significant reverse axial flux confined by a potential ambient field.  The necessary decrease of $f^2 (\psi )$ near the center of the flux rope causes the contour surfaces to crowd near the magnetic axis and the consequent strength of the non-axial flux makes deviations of the magnetic vector from the $y$-$z$ plane of more than $10^{\circ}$ very difficult to obtain.  One simple way to construct a model with reversing axial flux would be to include axial flux in the overlying bipolar arcade, of opposite sign to the axial flux in the flux rope.  A sheared flux rope emerging beneath an arcade of opposite shear may be the likeliest scenario for the creation of structures with reversing axial flux in the open solar corona.  We do not pursue this option here.

\subsection{Case in which temperature $T=T(\psi )$ is a flux function}
\label{tfluxfn}

We now turn our attention to non-force-free MHS solutions.  For the present we set $f^2 (\psi )$ to be as in Equation~(\ref{Pi1eq}) with $n=2$.  In the isothermal case, the hydrostatic scale height $\Lambda =p/(\rho g)$ is constant and we may set $\tau (z)=\exp (-z/\Lambda )$.  Now the condition $p/\rho =$~constant gives $p_1(\psi )=0$ and equation~(\ref{pintegral}) integrates to give

\begin{equation}
p=p_0 (\psi )\exp (-z/\Lambda ),
\end{equation}

\noindent
consistent with equation~(\ref{ideal}).

In the more general case with $T=T(\psi )$, Equation~(\ref{hydrostatic}) integrates to give

\begin{equation}
p(\psi ,z) = p_0(\psi )\exp \left(-\frac{z}{\Lambda (\psi )}\right) ,
\end{equation}

\noindent where $\Lambda (\psi )=k_BT(\psi )/(\mu g)$.  The Grad-Shafranov Equation~(\ref{reducedfb}) becomes

\begin{equation}
\nabla^2\psi +f(\psi )\frac{df(\psi )}{d\psi} +4\pi\left[\frac{d p_0(\psi)}{d \psi}+\frac{p_0(\psi )z}{\Lambda (\psi )^2}\frac{d\Lambda}{d\psi}\right]\exp\left( -\frac{z}{\Lambda (\psi )}\right) =0 .
\label{nonisogs}
\end{equation}

The isothermal case is $T(\psi )=$~constant or $\Lambda (\psi )=$~constant, which coincides with the case of Low~(1975).  The additional term for non-isothermal cases models the pressure gradient introduced by temperature changes across field lines.  In a system including very different temperatures, such as a cool prominence embedded in a much hotter coronal medium, this term may be important.

\begin{figure*}[ht]
\begin{center}
\resizebox{0.75\hsize}{!}{\includegraphics*{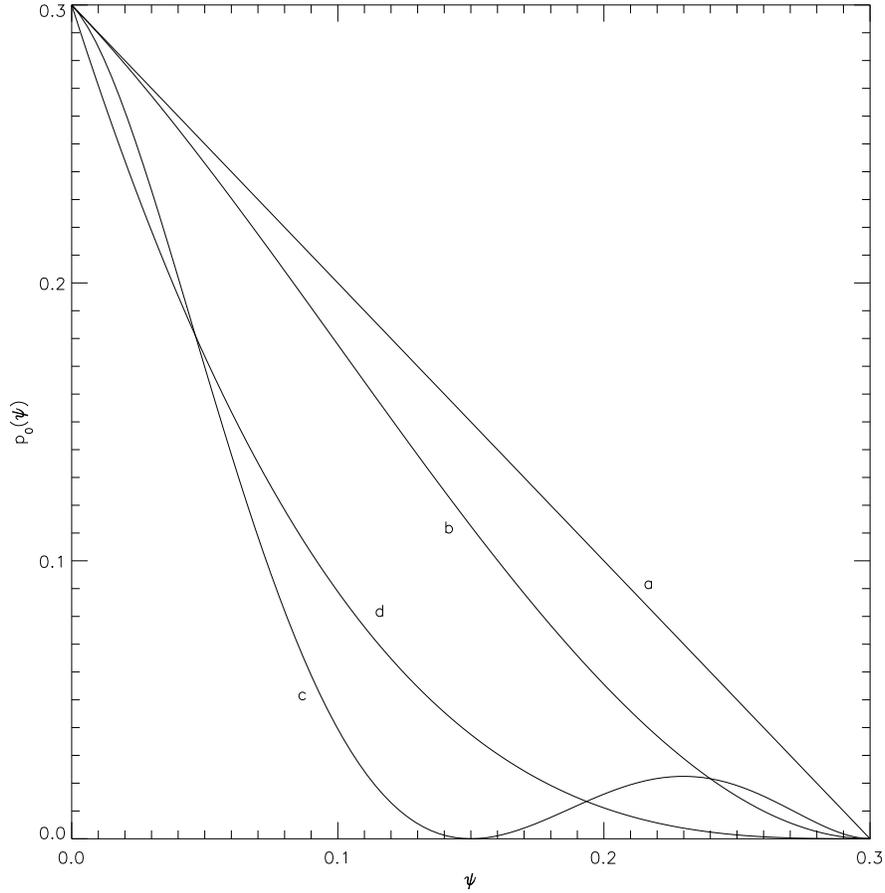}}
\end{center}
\caption{Examples of $p_0 (\psi )$ functions applied in this paper.  For $\psi >0.3$ the profiles are zero.}
\label{Pi2s}
\end{figure*}

The forms for $p_0 (\psi )$ considered in this work are plotted in Figure~\ref{Pi2s}.  Curve a is the form used by Low \& Hundhausen~(1995) and Zhang \& Low~(2004): $p_0 (\psi )=3/10 -\psi $.  For $\psi >3/10$, $p_0\equiv 0$ and the magnetic field is force-free there.  Curve b is the graph of

\begin{equation}
p_0 (\psi ) =\frac{1}{90}(3-10\psi )^2(3+10\psi ) .
\end{equation}

\begin{figure*}[ht]
\begin{center}
\resizebox{0.37\hsize}{!}{\includegraphics*{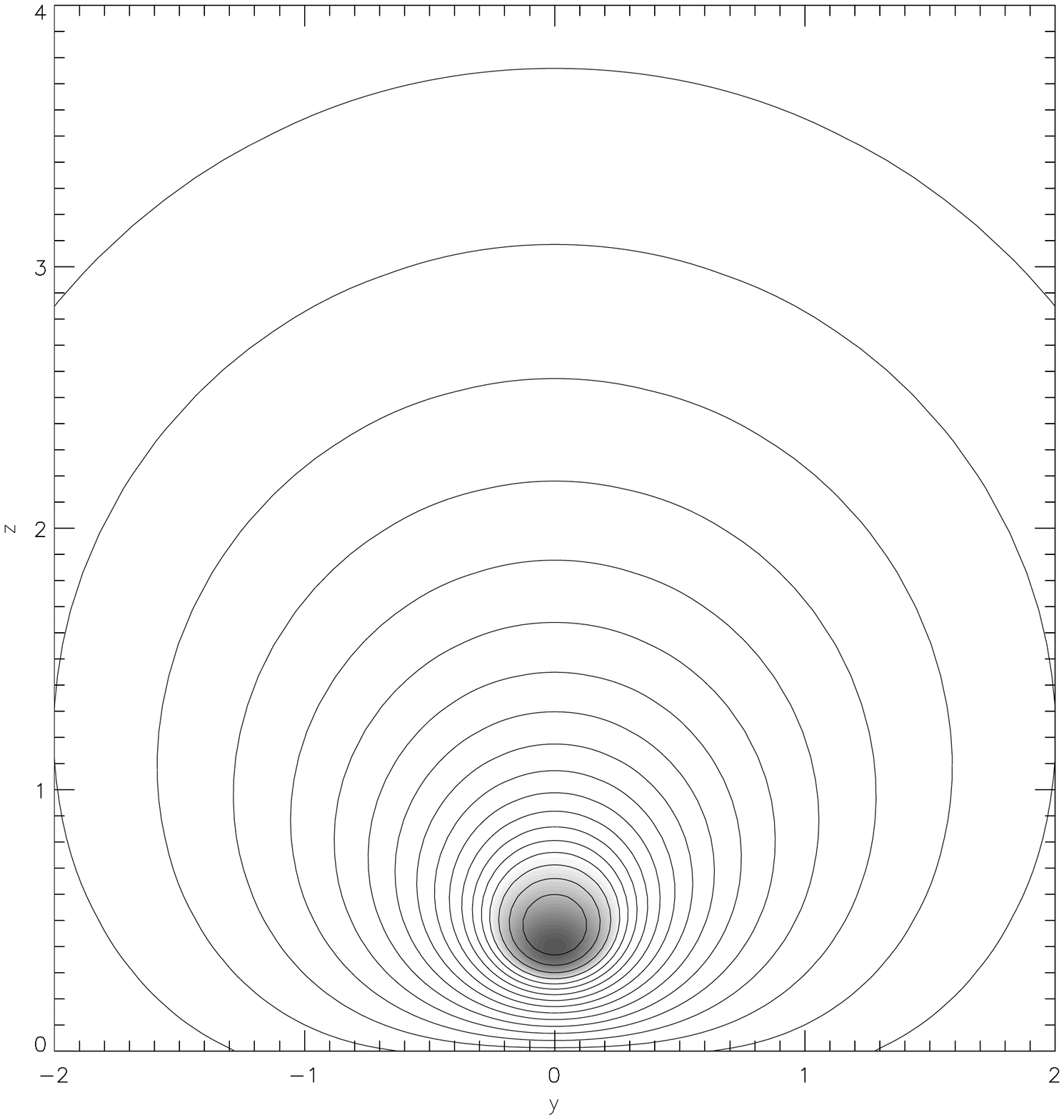}}
\resizebox{0.37\hsize}{!}{\includegraphics*{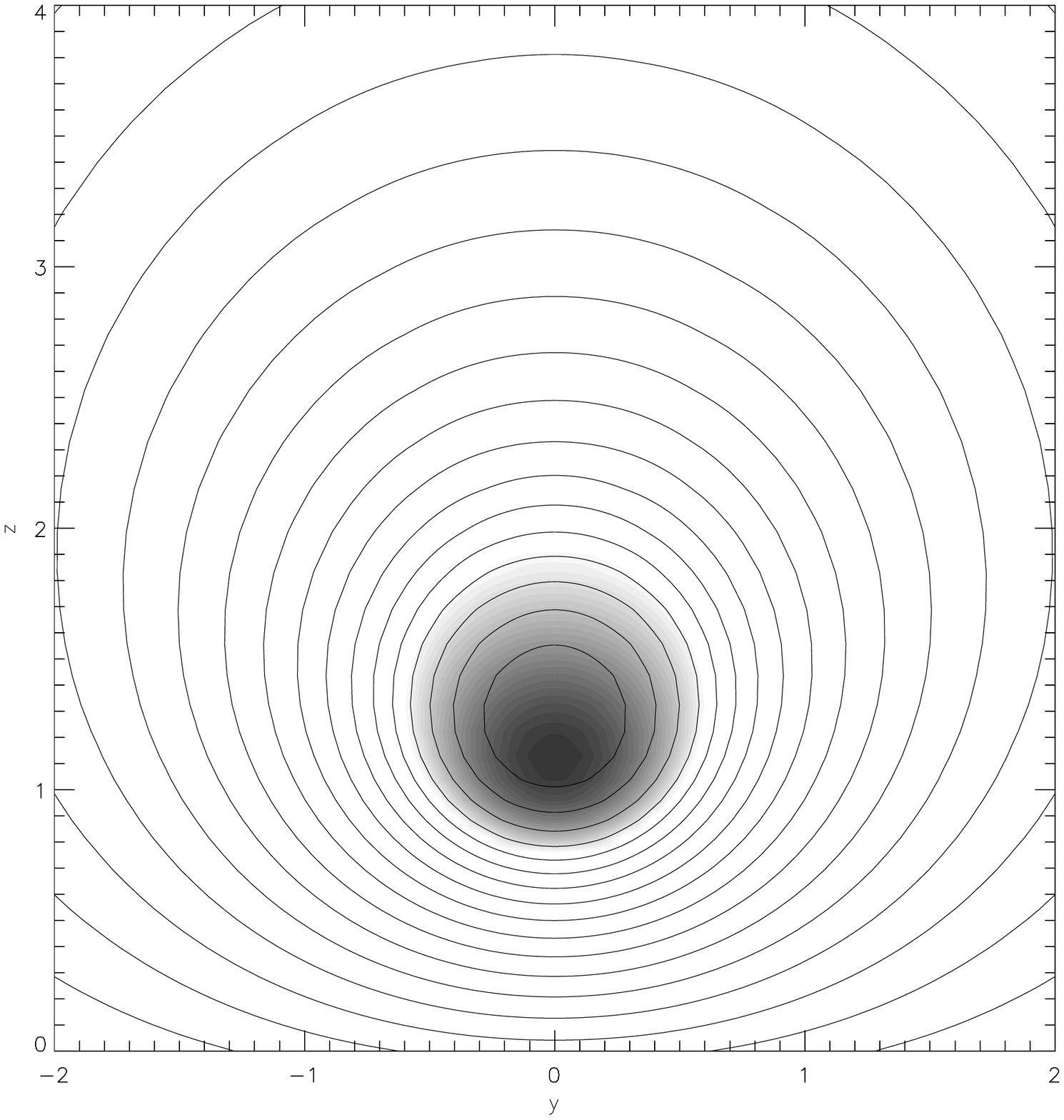}}
\resizebox{0.37\hsize}{!}{\includegraphics*{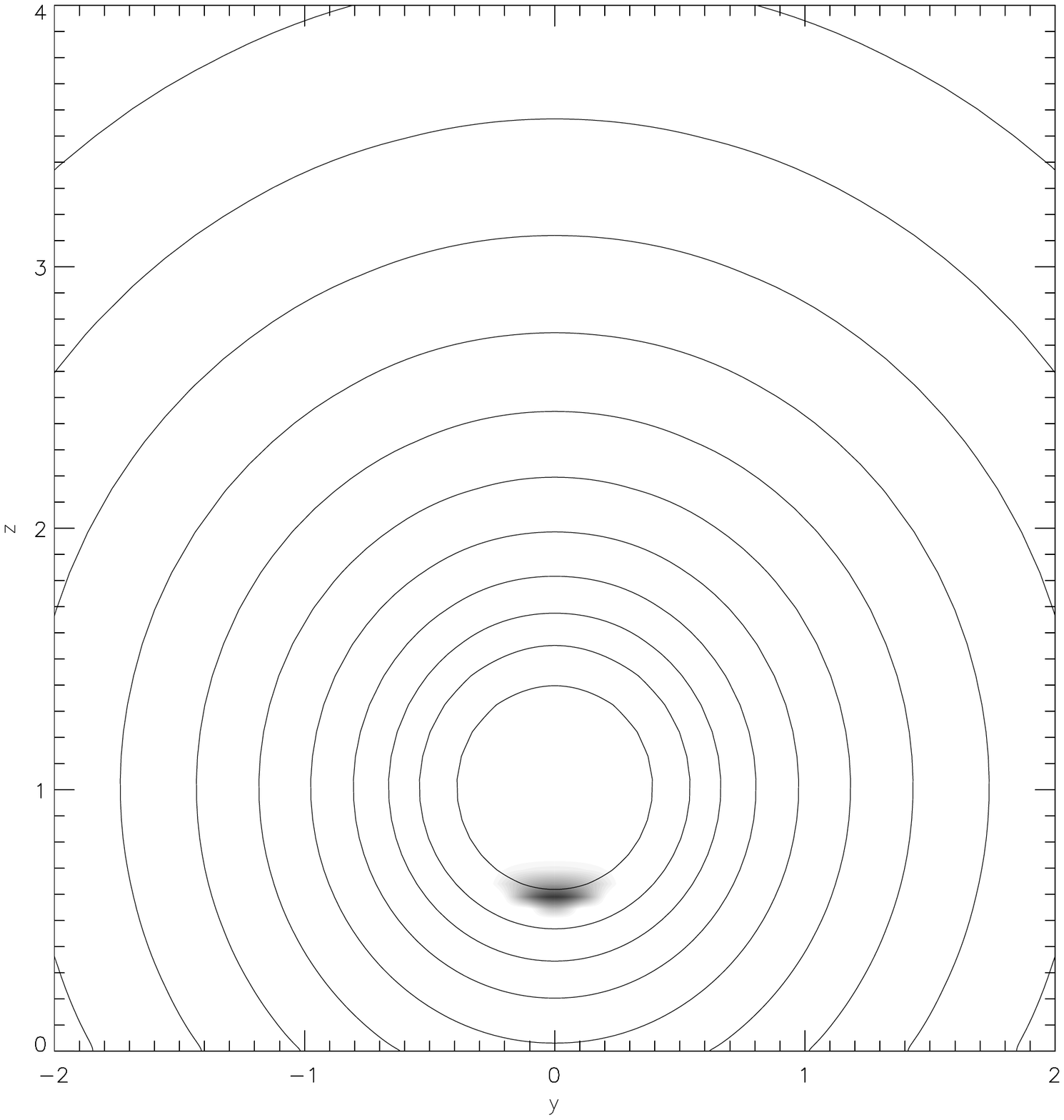}}
\resizebox{0.37\hsize}{!}{\includegraphics*{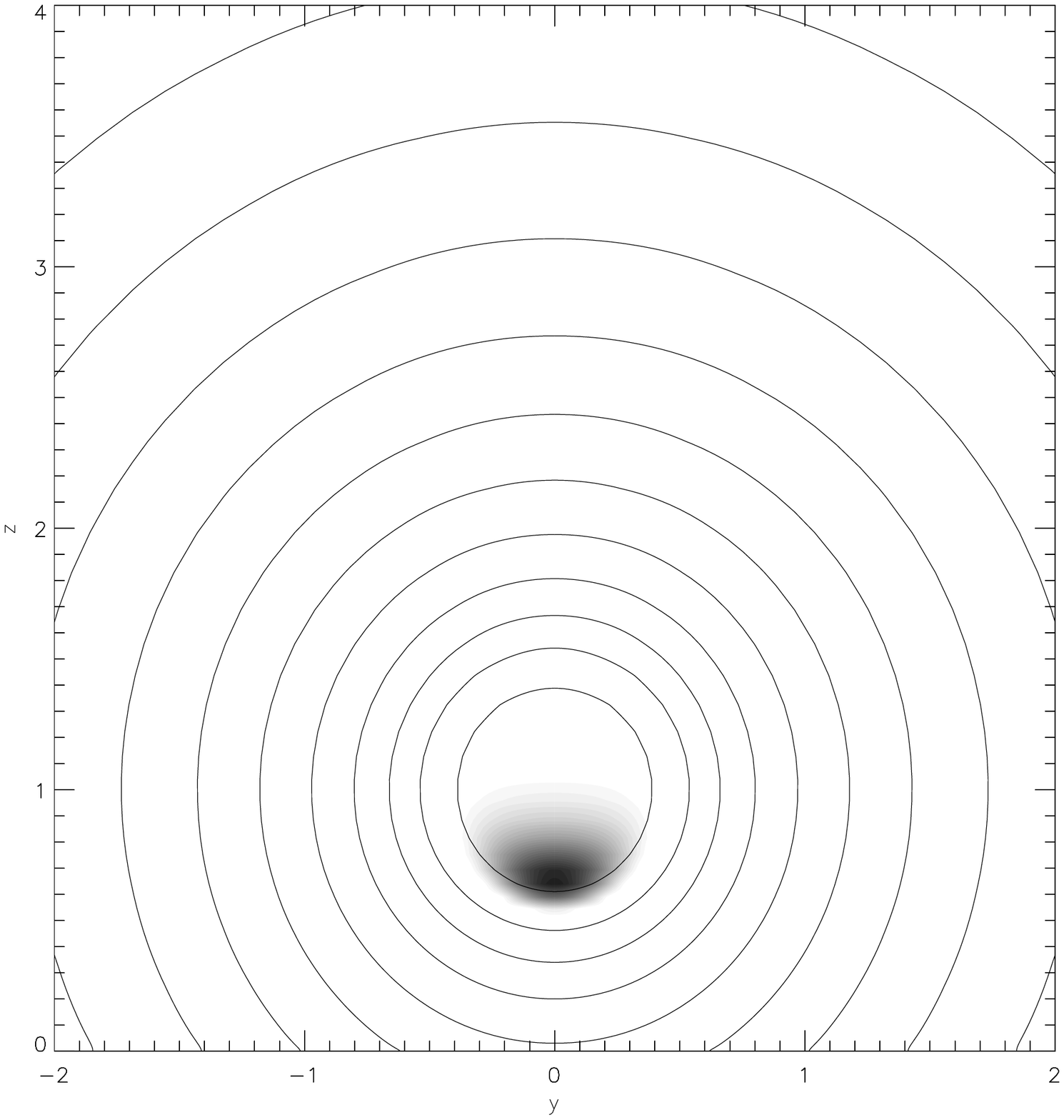}}
\end{center}
\caption{Magnetohydrostatic isothermal equilibria of inverse-polarity prominence magnetic fields: unsheared examples with $\Lambda =0.3$ (top left) and 1.0 (top right), and sheared examples with $\Lambda =0.03$ (bottom left) and 0.1 (bottom right).  The flux contours, which coincide with magnetic flux trajectories projected onto the $y$-$z$ plane, are represented by solid lines.  The plasma pressure is represented by the shading, with dark indicating high pressure.}
\label{tspot}
\end{figure*}

\noindent
Curve b is a smooth alternative to curve a, sharing its value and derivative at $\psi =0$ but having a double instead of single root at $\psi =3/10$.  We adopt this form of $p_0 (\psi )$ for the examples plotted in Figure~\ref{tspot}.

The examples shown in the top two panels of Figure~\ref{tspot} are solutions to the same equation with the same boundary conditions but with different values for the temperature.  These first two examples have no axial magnetic flux and have smaller plasma $\beta =8\pi p/|{\bf B}|^2$ than the sheared examples plotted below them.  The examples of the bottom pictures the plasma $\beta$ has maximum of about 1/10 or 1/5.  The $\beta$ parameter is infinite at the unsheared solutions' neutral points at the center of the flux ropes, but a typical value for $\beta$ in these solutions is about 10.  The Bernoulli integral along field lines is

\begin{equation}
\int\frac{dp}{\rho} +gz = \Lambda g\log\left(\frac{\rho}{\rho_0}\right) +gz ={\rm constant},
\end{equation}

\noindent where $\rho_0$ is the density at $z=0$.  This expresses the fact that the thermal energy and gravitational potential energy must add to give a conserved quantity, the total energy.  Clearly the thermal energy is directly proportional to the hydrostatic scale height and therefore the temperature.  The top pictures of Figure~\ref{tspot} show that an increase of thermal energy of the plasma causes the thermal pressure to be more effective in pushing out against the confining forces of the magnetic field.  The more even spread of mass across the concave-upward floor of the magnetic configuration imposes less stress on the magnetic field at $y=0$ and allows the prominence plasma to sit higher in the atmosphere than in cases with lower temperature.  The effect of this on the characteristic scale length of both the plasma distribution and the magnetic structure itself are clearly visible in the top two pictures.  

The bottom two pictures show examples with colder plasma but with significant axial magnetic flux density - they have axial flux profiles derived from that of model b of Section~\ref{nlff} while the magnitudes of their $p_0$ profiles are similar to those of the top pictures.  These stronger magnetic structures are able to accommodate cold plasma without significant changes in their magnetic fields. In reality prominences have typical hydrostatic scale height of a few hundred km while flux ropes are of order 10~Mm across.  The bottom left picture shows the model closest to these parameter values.  

To demonstrate the flexibility of the solution method, we show an example with two separate prominence concentrations in a single flux rope.  Curve c in Figure~\ref{Pi2s} is,

\begin{equation}
p_0 (\psi )=\frac{200000}{81}\left( \frac{3}{10} -\psi \right)^2\left( \frac{3}{20} -\psi\right)^2\left( \frac{3}{50}+\psi \right) .
\end{equation}

\begin{figure*}[ht]
\begin{center}
\resizebox{0.37\hsize}{!}{\includegraphics*{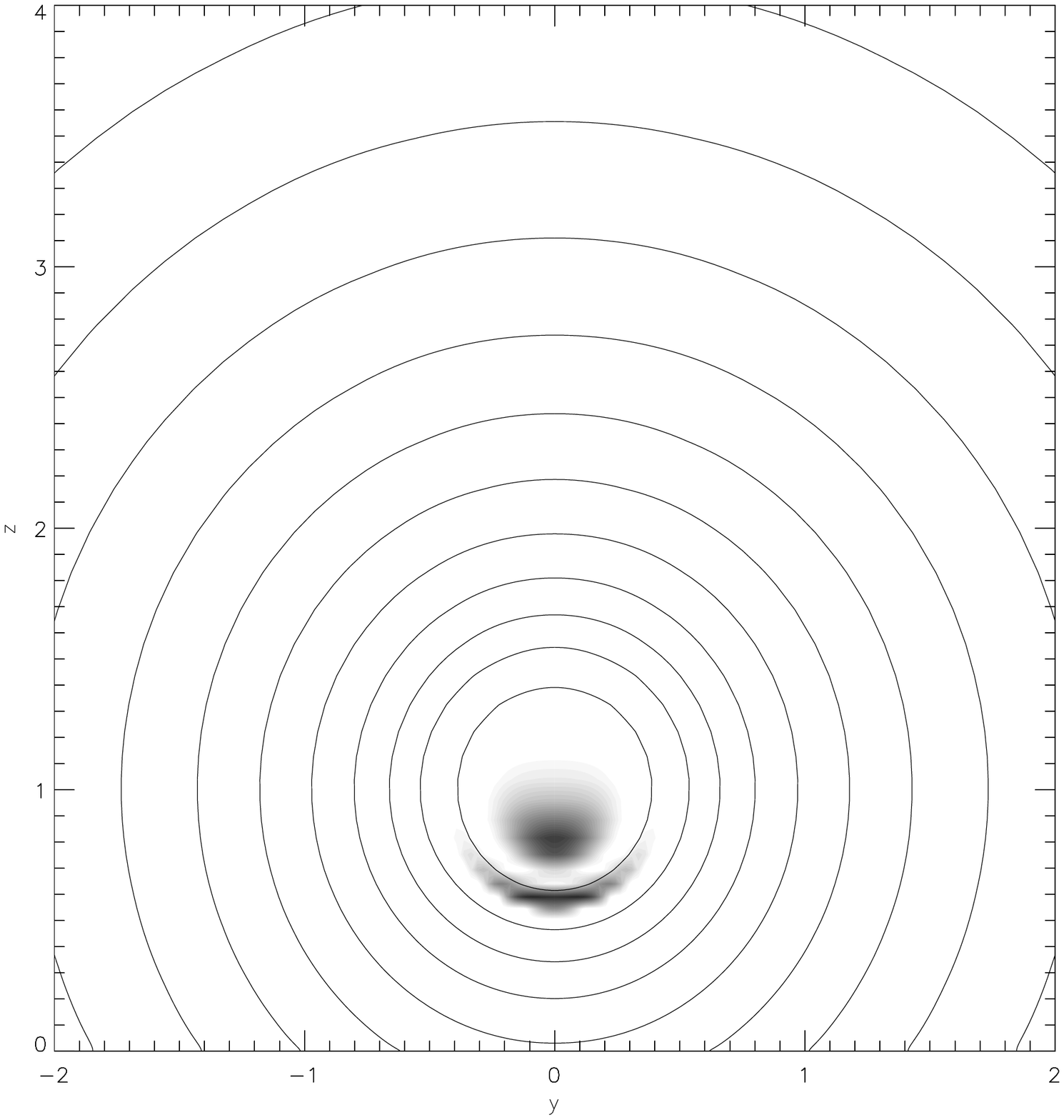}}
\end{center}
\caption{The flux contours and plasma pressure of a magnetohydrostatic isothermal equilibrium with an inverse-polarity prominence magnetic flux rope containing a double prominence enhancement.}
\label{double}
\end{figure*}

\noindent
This polynomial has double roots at $\psi =3/10$ and $3/20$ and has the same value and derivative at $\psi =0$ as $3/10 -\psi$.  The corresponding prominence model is plotted in Figure~\ref{double}.  A narrow evacuated channel, corresponding to the first zero of $p_0 (\psi )$, separates a rounded blob of plasma above and a curved sheet below.

\begin{figure*}[ht]
\begin{center}
\resizebox{0.37\hsize}{!}{\includegraphics*{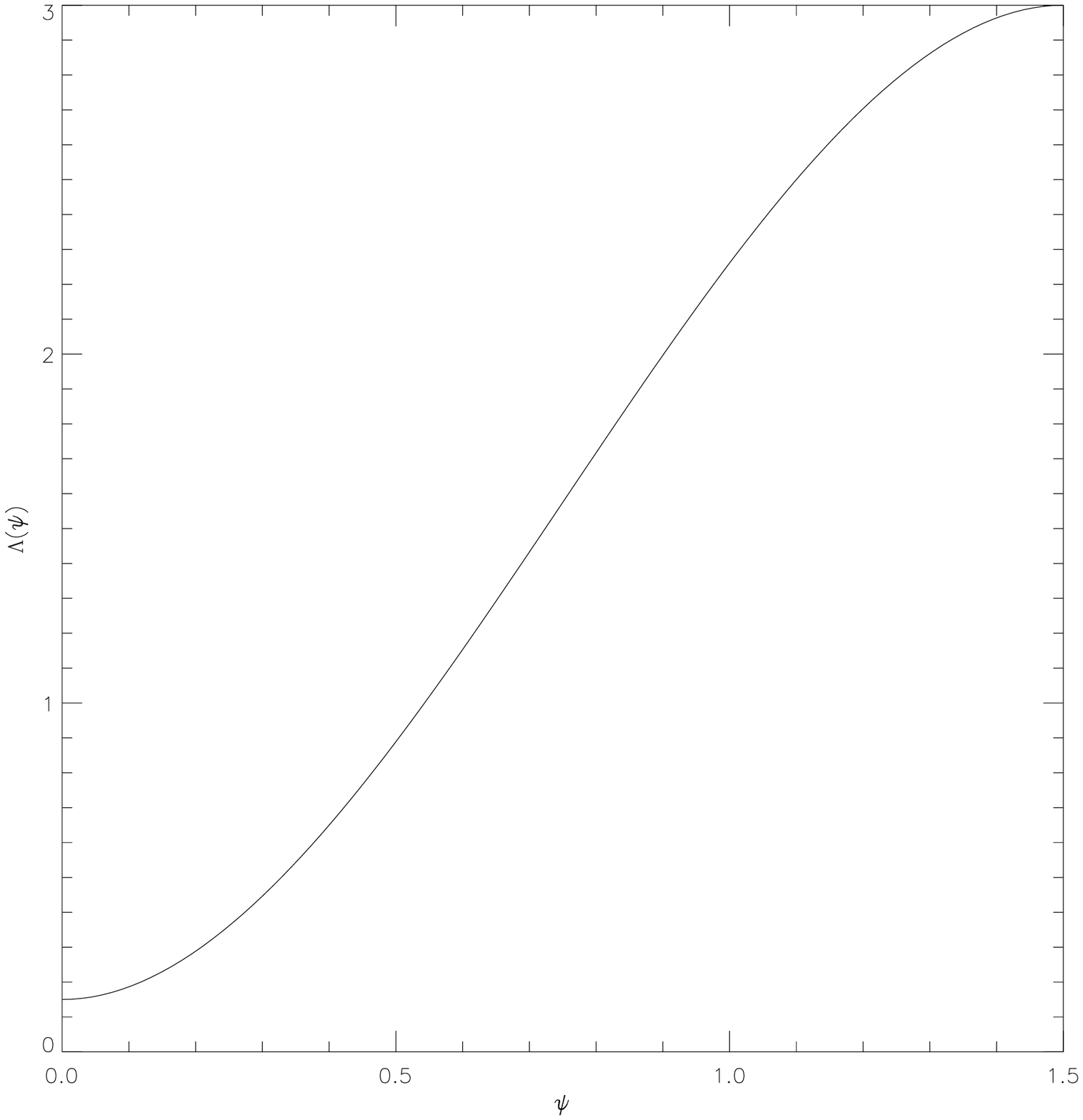}}
\resizebox{0.37\hsize}{!}{\includegraphics*{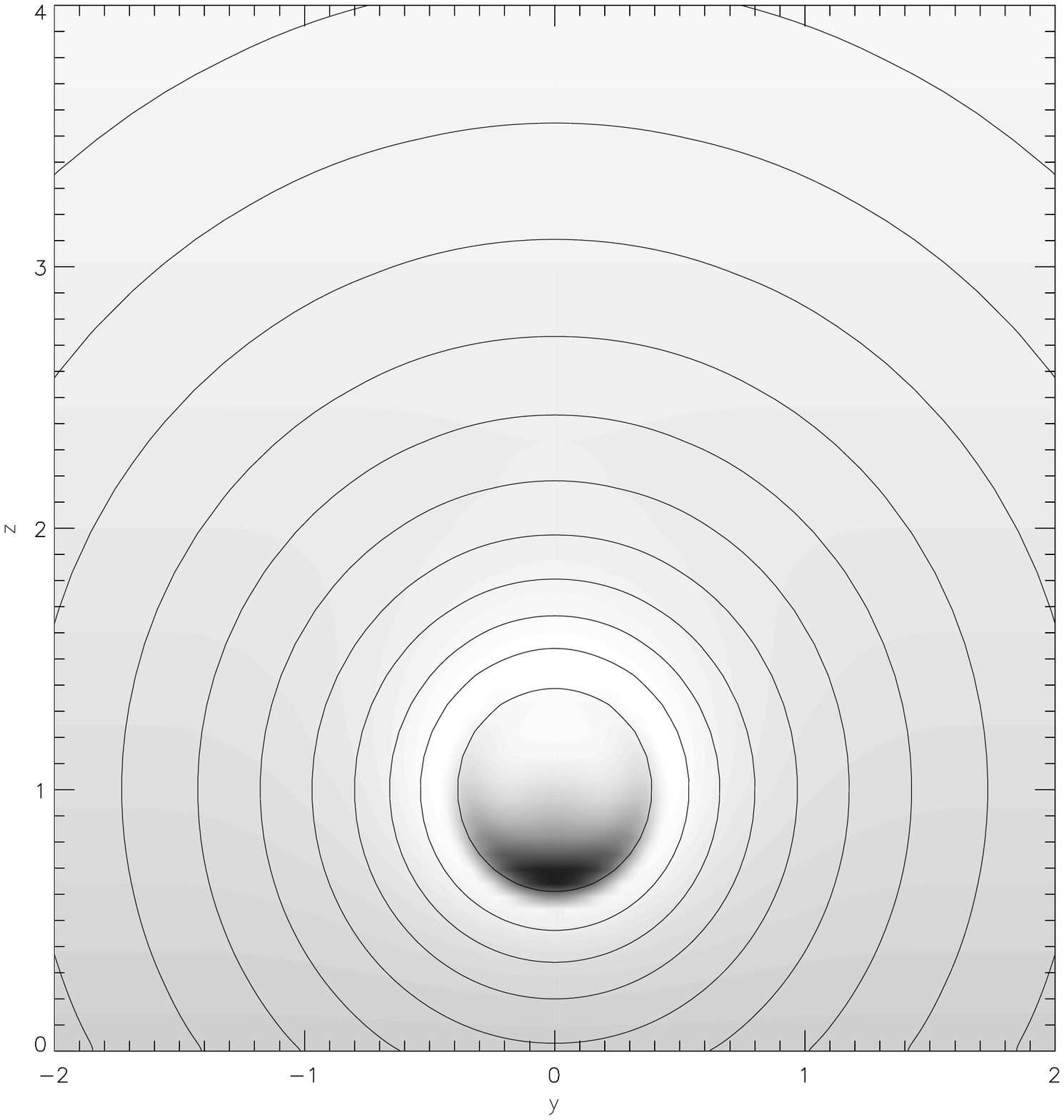}}
\end{center}
\caption{A cool, dense prominence plasma enhancement suspended in a hot, sparse corona.  The left picture shows a highly non-isothermal scale height profile.  The right picture shows the flux contours and the plasma pressure.  Note the evacuated cavity within which the dense plasma is embedded.}
\label{noniso}
\end{figure*}

Further complexity is introduced by an ambient atmosphere populated by a hot plasma in a non-force-free magnetic field.  Figure~\ref{noniso} shows a model which exploits the non-isothermal aspect of Equation~(\ref{nonisogs}).  A $\psi$-dependent temperature profile is imposed so that a hot corona and a cool, dense prominence can be modeled together in a self-consistent way.  The variation of the hydrostatic scale height $\Lambda (\psi )$, proportional to the temperature,  with $\psi$ within the flux contour $\psi =3/2$ is described by

\begin{equation}
\Lambda (\psi )=\frac{3}{20}+\frac{19}{5}\psi^2\left(1-\frac{4}{9}\psi\right) .
\end{equation}

This profile is graphed in the left panel of Figure~\ref{noniso}.  The scale height therefore ranges from $3/20$ to $3$ between the magnetic axis and $\psi =3/2$.  Outside $\psi =3/2$ the temperature has a constant value of $3$.  On the other hand, the plasma pressure is controlled by choosing $p_0(\psi )$, which in this model is based on curve d of Figure~\ref{Pi2s}.  Curve d is similar to curve b but has derivative -3 at $\psi =0$.  The equation for curve d is

\begin{equation}
p_0 (\psi )=\frac{1}{90}(3-10\psi )^3 ,
\end{equation}

\noindent
so that this curve has a triple root at $\psi =3/10$.  This $p_0(\psi )$ is zero outside the contour $\psi =0.3$.  In order to generate an ambient atmosphere we add a small number, 0.0005, to $p_0(\psi )$ in all space.  Therefore, outside a certain contour, $\psi =1.5$, the pressure contours are planes parallel to the photosphere.  Within the contour $\psi =3/10$ is a cool, dense prominence structure gravitationally stratified with scale height about 1/20 of that in the ambient corona.  One result of this difference in stratification between the interior and the exterior of the flux rope is that the prominence plasma resides in a cavity: the portion of the flux rope not occupied by dense prominence plasma is more sparsely populated even than the corona.  This is an essential part of the three-part structure often observed in prominences and CMEs (Low \& Hundhausen~1995, Gibson et al.~2006).  Low \& Hundhausen~(1995)  emphasized and gave physical meaning to the three-part structure of the prominence-flux rope-helmet system often observed.  For simplicity, the prominence is usually studied without accounting for the low-density cavity in which it is observed to be embedded.  While the large-scale corona varies in structural complexity with the solar cycle, the basic three-part helmet structure seems to persist throughout this cycle.  Here it is seen as a natural feature of a low-density ambient atmosphere containing a flux rope with a temperature minimum and a plasma pressure maximum at the flux rope center.

\subsection{Case in which entropy $s=p/\rho^{\gamma}=s(\psi )$ is a flux function}
\label{efluxfn}

In the polytropic case, Equation~(\ref{hydrostatic}) takes the form

\begin{equation}
\frac{\gamma}{\gamma -1} \left.\frac{\partial\rho^{\gamma -1}}{\partial z}\right|_{\psi={\rm const}} = -\frac{g}{s(\psi )} ,
\end{equation}

\noindent where $p(\psi ,z)=s(\psi )\rho(\psi ,z)^{\gamma}$.  In the special case of Low \& Zhang~(2004), $p_0(\psi )=\rho_0 (\psi )g/n$ for $n=\gamma /(1-\gamma )$.

\noindent The Bernoulli integral along field lines is

\begin{equation}
\int\frac{dp}{\rho} +gz = \frac{\gamma}{\gamma-1}\left(\frac{p}{\rho} -\left.\frac{p}{\rho}\right|_{z=0}\right) +gz ={\rm constant},
\end{equation}

\noindent expressing the fact that the thermal energy and gravitational potential energy must add to give a conserved quantity.  For $\gamma < 1$, $p/\rho -(p/\rho)|_{z=0}$ must be $> 0$, i.e. the temperature increases with height while for $\gamma > 1$, $p/\rho -(p/\rho)|_{z=0}$ must be $< 0$, i.e. the temperature decreases with height.  Note that along field lines the temperature is a linear function of height, which must therefore cross zero and become negative at some height.  Such solutions may still be used to model solar prominences since they are of finite vertical extent.  In the separable case, the plasma pressure and temperature are given by

\begin{eqnarray}
p (\psi ,z) & = & p_0(\psi )\left[-\frac{\gamma -1}{\gamma}\frac{(z-z_0)}{\lambda}\right]^{\frac{\gamma}{\gamma-1}},\\
T & = & -\frac{\mu}{k_B} \frac{(\gamma -1)}{\gamma} g(z-z_0),
\end{eqnarray}

\noindent where $\lambda =p_0(\psi )/\rho_0 (\psi)g$ is a constant length scale.  The Grad-Shafranov Equation~(\ref{reducedfb}) is

\begin{equation}
\nabla^2\psi +f(\psi )\frac{df(\psi )}{d\psi} +4\pi\left[-\frac{\gamma -1}{\gamma}\frac{(z-z_0)}{\lambda}\right]^{\frac{\gamma}{\gamma-1}}\frac{dp_0(\psi )}{d\psi} =0 .
\end{equation}

\noindent Note that the temperature is independent of $\psi$ in this separable case, since $p$ and $\rho$ have common $\psi$-dependence, and the position of this zero-point is $z=z_0$ on all field lines.  In order for the plasma pressure and temperature to be well-defined and positive in the domain of interest, the factor $-(\gamma-1)(z-z_0)$ must be positive throughout that domain.  The gradient of the linear variation of $T$ with altitude is determined entirely by the polytropic index $\gamma$.  $T$ is an increasing function of height for $\gamma < 1$ and a decreasing function of height for $\gamma > 1$.  Thus, for example, Low \& Zhang chose their $z_0$ to be negative to keep this critical point out of the way beneath the base of the corona because they were working with values of $\gamma <1$.  For cases with $\gamma <1$, $z_0$ must be placed above the domain of interest and for $\gamma>1$ $z_0$ must be placed beneath.  If $\gamma >1$ this introduces difficulties for the construction of solutions in which all of space is populated by plasma, such as the one in Figure~\ref{noniso}, but for isolated prominence plasma enhancements in empty atmospheres these solutions are appropriate.

For the general case $\rho_0 =\rho_0 (\psi )$, the solution is not separable and the pressure and temperature are given by

\begin{eqnarray}
p (\psi ,z) & = & p_0 (\psi ) \left[ 1-\left(\frac{\gamma -1}{\gamma}\right)\frac{(z-z_0)}{\Lambda (\psi )}\right]^{\frac{\gamma}{\gamma-1}},\\
T & = & -\frac{\mu}{k_B} \frac{(\gamma -1)}{\gamma} g(z-z_0) +T_0(\psi )\nonumber\\
& = & -\frac{\mu}{k_B} \frac{(\gamma -1)}{\gamma} g\left[z-\left( z_0+\frac{\gamma}{\gamma-1}\Lambda (\psi )\right)\right],
\end{eqnarray}

\noindent where $$T_0(\psi )=\frac{\mu}{k_B} \frac{p_0(\psi )}{\rho_0 (\psi )}$$ is the temperature as a function of $\psi$ at $z=z_0$, and $\Lambda (\psi )=p_0(\psi )/\rho_0 (\psi)g$ is the hydrostatic scale height at $z=z_0$ on each flux surface labeled $\psi$.  These are both zero in the separable case discussed above.  The non-separable Grad-Shafranov Equation~(\ref{reducedfb}) is

\begin{eqnarray}
\lefteqn{ \nabla^2\psi +f(\psi )\frac{df(\psi )}{d\psi} +4\pi\left[1-\left(\frac{\gamma -1}{\gamma}\right)\frac{(z-z_0)}{\Lambda (\psi )}\right]^{\frac{1}{\gamma-1}}\times } \nonumber\\
& & \left[\left(1-\frac{\gamma -1}{\gamma}\frac{(z-z_0)}{\Lambda (\psi )}\right)\frac{d p_0(\psi )}{d\psi} -\frac{p_0(\psi )(z-z_0)}{\Lambda(\psi )^2}\frac{d \Lambda(\psi )}{d\psi}\right] =0 .
\end{eqnarray}

\noindent In this general case $T$ still varies linearly with $z$ on each individual field line with gradient unaffected by $T_0(\psi )$, but that $T$ can now vary across field lines via $T_0(\psi )$.  The effect of $T_0 (\psi)$ is therefore simply to shift the critical point $z=z_0^{'}=z_0+[\gamma/(\gamma-1)]\Lambda (\psi )$ vertically by a distance $[\gamma/(\gamma-1)]\Lambda (\psi )$ on each field line.  This quantity can vary freely from field line to field line but its physical effect is indistinguishable from the constant critical point of the separable case $z=z_0$: fixing the temperature to be a given temperature at a given height fixes $z_0^{'}$ exactly as $z_0$ is fixed in the separable case.

\begin{figure*}[ht]
\begin{center}
\resizebox{0.37\hsize}{!}{\includegraphics*{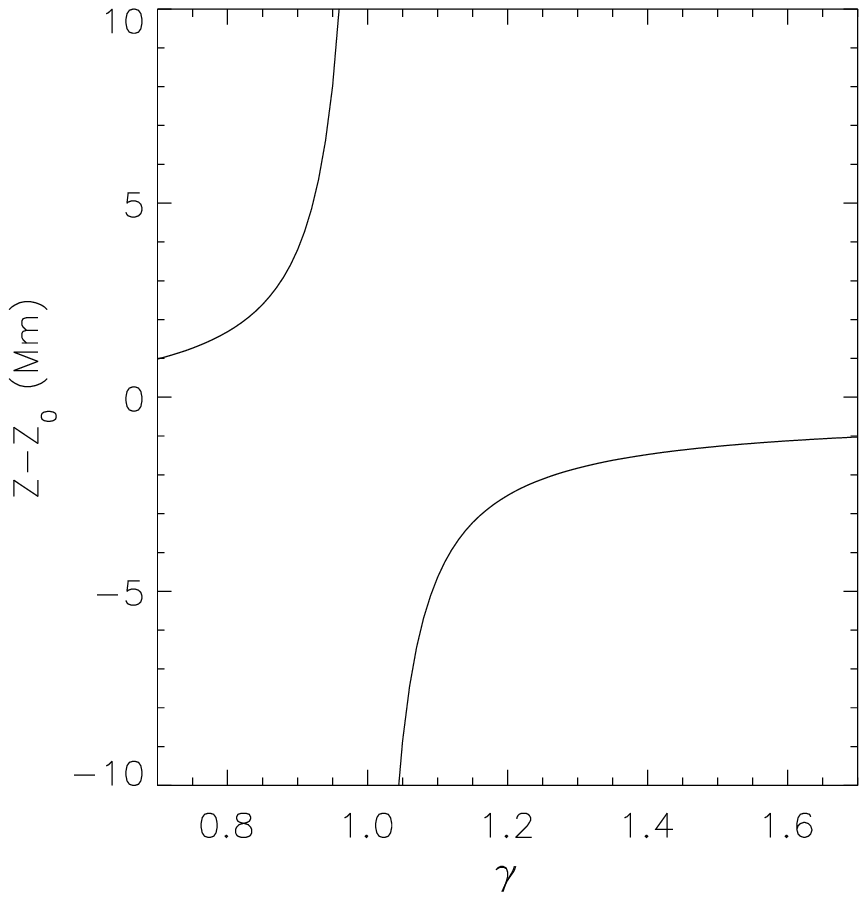}}
\resizebox{0.37\hsize}{!}{\includegraphics*{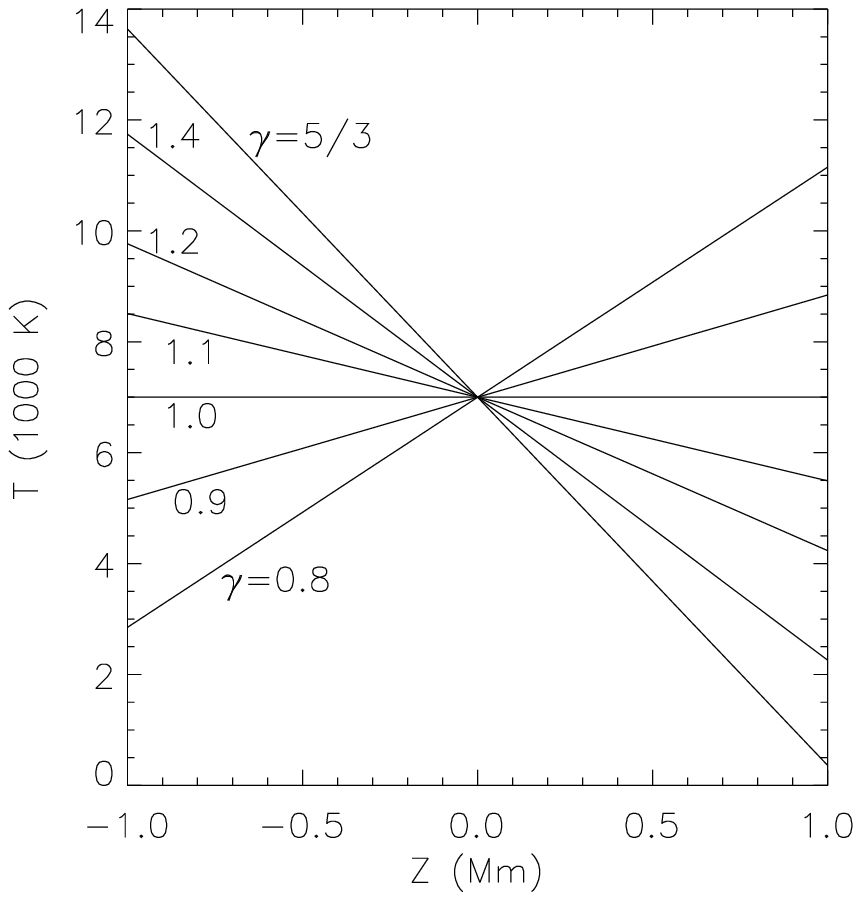}}
\resizebox{0.37\hsize}{!}{\includegraphics*{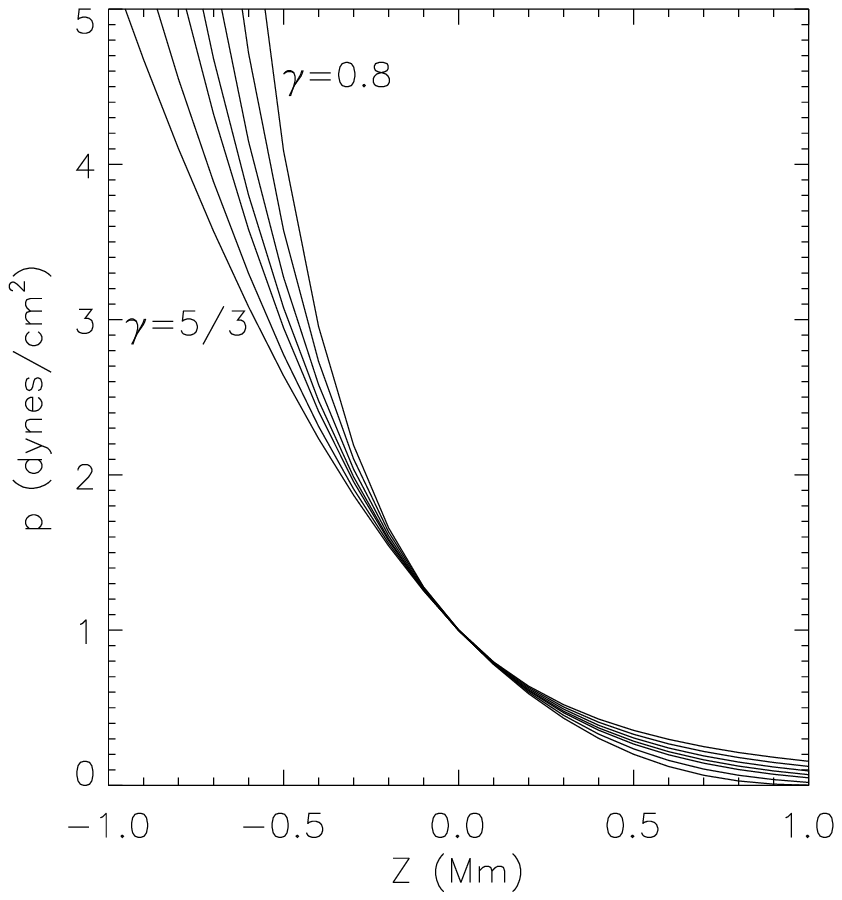}}
\resizebox{0.37\hsize}{!}{\includegraphics*{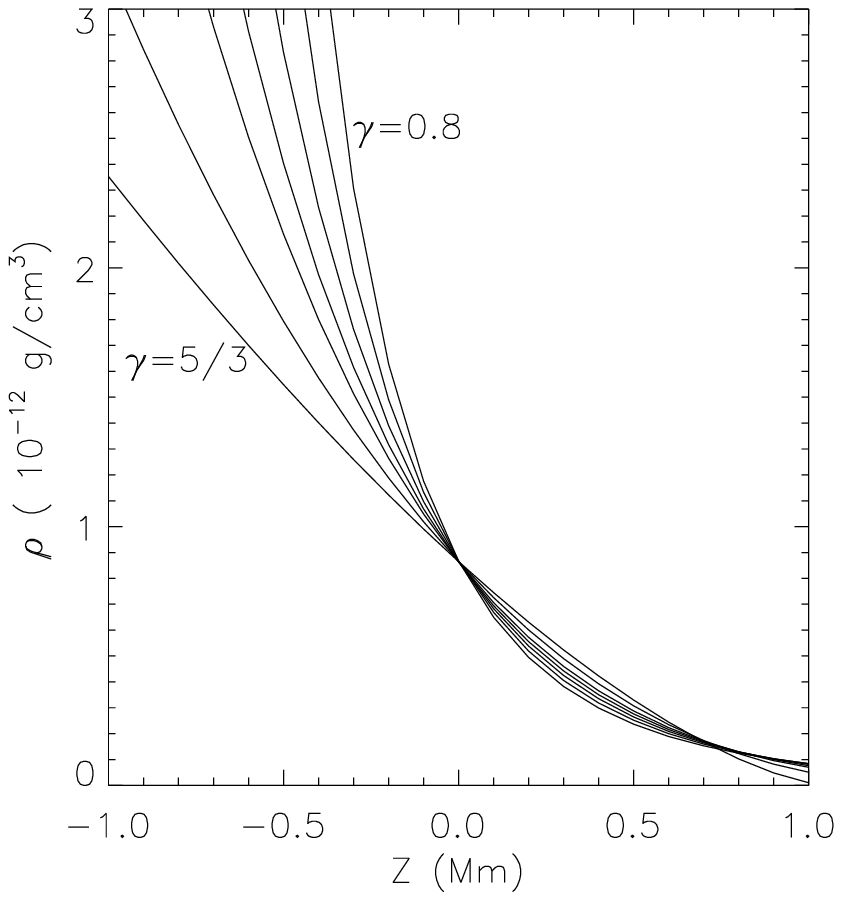}}
\end{center}
\caption{Hydrostatic 1D polytropic equilibrium: for a fixed temperature at $z=0$, the position of the critical point $z=z_0$ is graphed as a function of the polytropic index $\gamma$ (top left).  The resulting temperature (top right), pressure (bottom left) and density (bottom right) are shown as functions of height.}
\label{polyhydrostatic}
\end{figure*}

Some illustrative polytropic hydrostatic 1D solutions are shown in Figure~\ref{polyhydrostatic}.  The location of a point $z$ at a particular chosen temperature (in this case 7000~K, a typical temperature for prominence plasma) relative to the critical point $z_0$ is graphed as a function of $\gamma$ in the top left picture.  As anticipated earlier, for $\gamma </>1$, $z_0$ is below/above the domain of interest.  Since the temperature gradient is larger for values of $\gamma$ further from 1, the critical point must be positioned closer to the domain of interest for these values of $\gamma$ and far from the domain of interest for $\gamma$ close to 1.  In the limiting case $\gamma =1$, the critical point asymptotes, crossing from $-\infty$ to $\infty$.

In Figure~\ref{polyhydrostatic} the plasma parameters $p$, $\rho$ and $T$ for different values of $\gamma$ are forced to be equal at $z=0$.  Moreover, from the balance of pressure gradient and weight, the vertical derivatives of the pressures are also equal at $z=0$ so that all pressure curves are tangential there.  Among the pressure curves is the $\gamma =1$ isothermal curve which has constant pressure scale height as discussed in the last subsection.  Solutions with $\gamma >1$ are cooler than the $\gamma =1$ solution in $z>0$, have smaller scale heights and therefore have a steeper appearance there.  On the other hand, solutions with $\gamma <1$ are hotter than the $\gamma =1$ solution in $z>0$, have larger scale heights and therefore have a flatter appearance.  Meanwhile, in $z<0$ the opposite is true: solutions with $\gamma >/<1$ are hotter/cooler and have flatter/steeper pressure curves.

The density gradients, on the other hand, are not equal at $z=0$ since they are affected by the different temperature gradients of the different solutions.  Density gradients are larger for small $\gamma$ values than for larger $\gamma$ values since the plasma is cooler.  They are also larger within a scale height ($\approx 400$~km) of $z=0$ in the positive direction, where the temperature gradient causes the density to fall off more quickly for $\gamma <1$ and more slowly for $\gamma >1$ than in the $\gamma =1$ case.  Beyond a scale height, the higher temperature of the $\gamma <1$ plasma causes these density curves to become flatter than the $\gamma =1$ curve and the lower temperature of the $\gamma >1$ plasma causes the density curves to become steeper than the $\gamma =1$ curve.

\begin{figure*}[ht]
\begin{center}
\resizebox{0.37\hsize}{!}{\includegraphics*{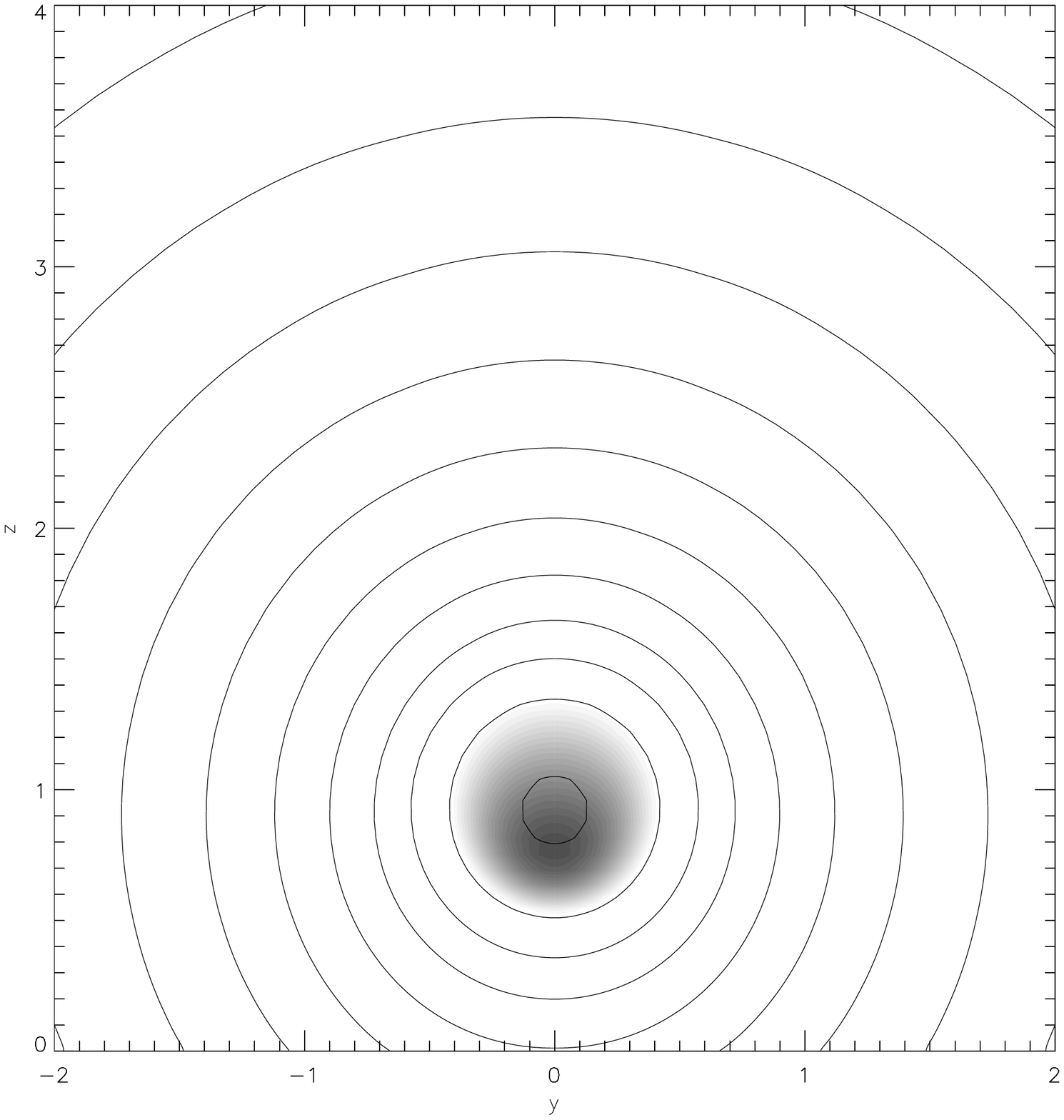}}
\resizebox{0.37\hsize}{!}{\includegraphics*{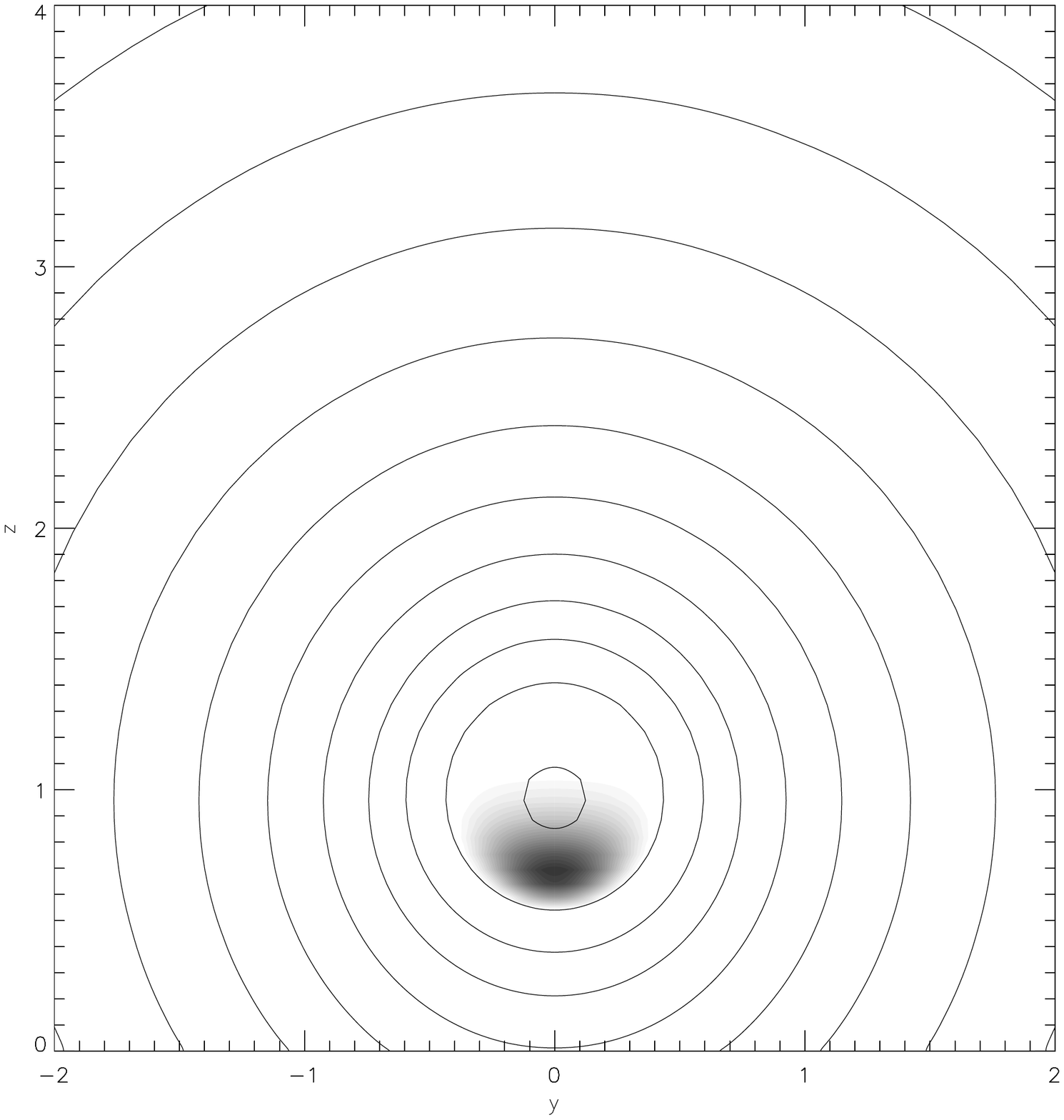}}
\resizebox{0.37\hsize}{!}{\includegraphics*{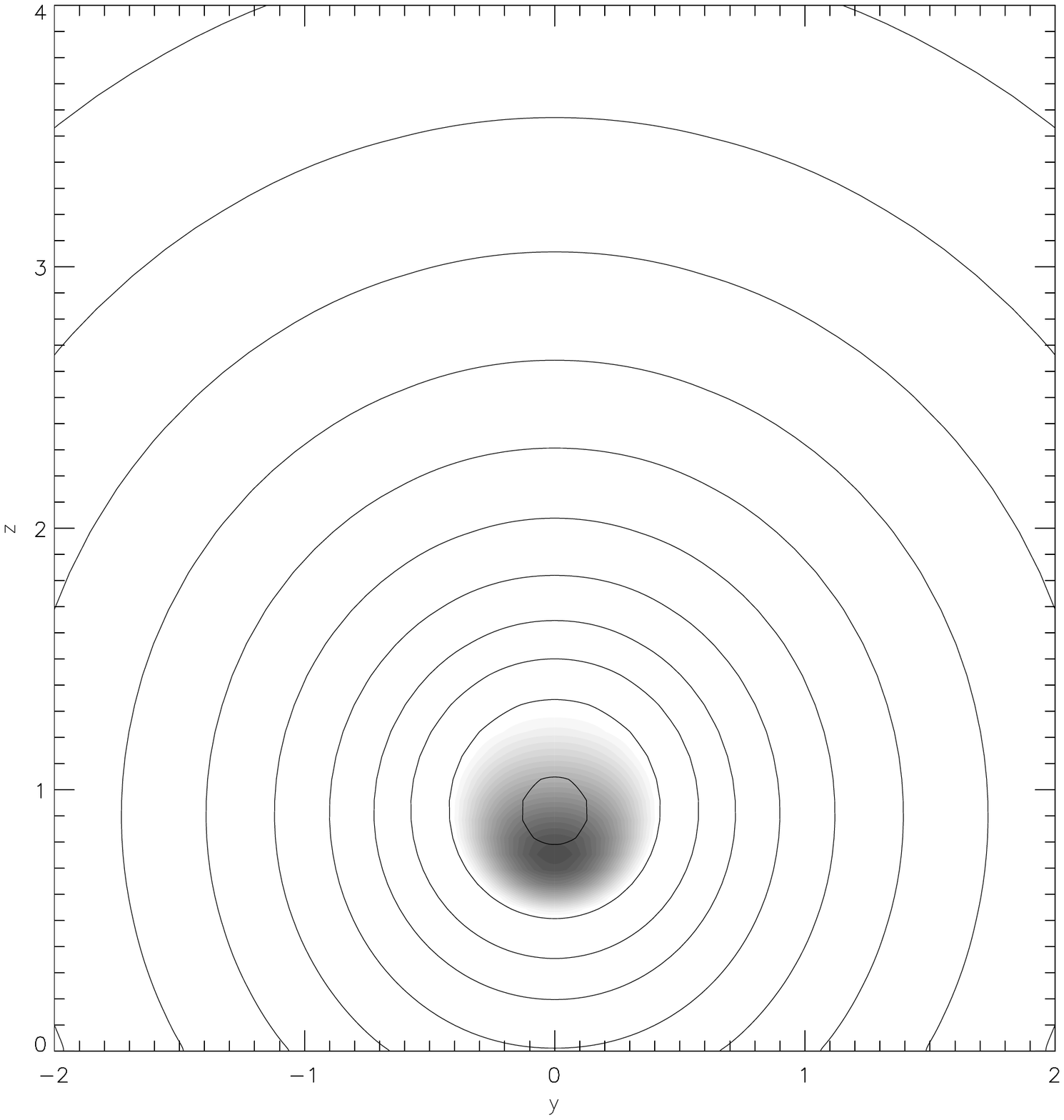}}
\resizebox{0.37\hsize}{!}{\includegraphics*{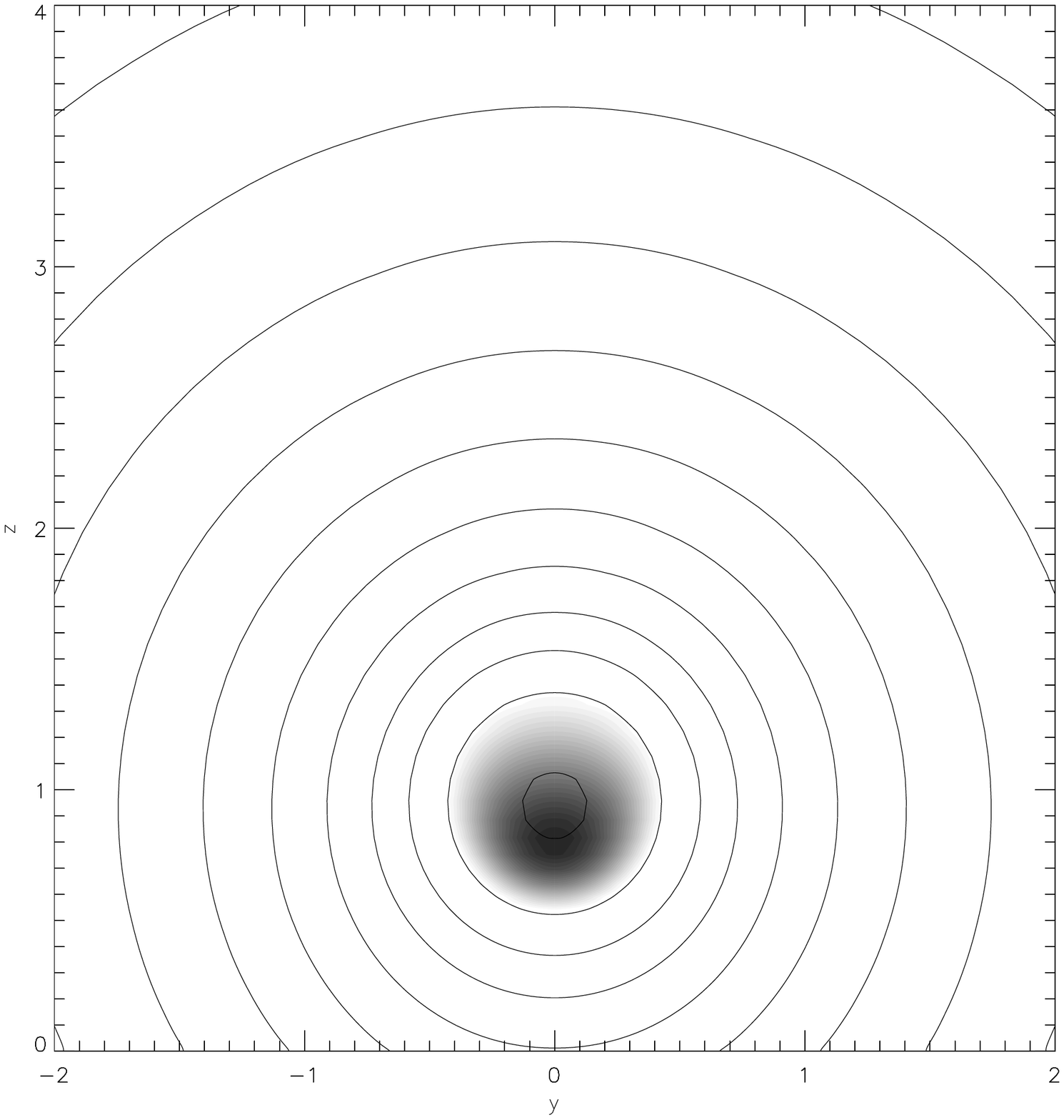}}
\end{center}
\caption{Magnetohydrostatic polytropic equilibrium: $\gamma =0.8$ (top left), 1.1 (top right), 1.4 (bottom left) and 5/3 (bottom right).  The critical point $z_0$ is located at $z=0$ for the $\gamma =0.8$ case and at $z=2$ for the $\gamma >1$ cases.}
\label{largegamma}
\end{figure*}

Figure~\ref{largegamma} shows magnetohydrostatic 2D examples with a range of values of $\gamma$.  The 1D hydrostatic solutions of Figure~\ref{polyhydrostatic} and the 2D magnetohydrostatic solutions here are related in that the 2D solutions restricted to flux contours $\psi =$~constant are 1D hydrostatic solutions of the type described above.  The 2D solutions were all derived with the same boundary conditions as those presented in earlier sections and the differences between the plots in Figure~\ref{largegamma} are due only to the different values of $\gamma$ and the locations of $z_0$.  In the $\gamma =0.8$ case $z_0$ must be below the flux rope while in the cases with $\gamma >1$ it must be above.  In the four examples shown, $z_0$ is placed about a unit from the center of the flux rope.  This means that for $\gamma >1$, the temperature increases as $\gamma $ increases, as can be seen in the plots.  In these cases the hotter plasma is at the bottom of the flux rope while the hotter plasma is at the top in the $\gamma <1$ case.  This makes the $\gamma >1$ plasma pressure distributions more bottom-heavy than the $\gamma <1$ case.


\subsection{A prominence with normal topology}

\begin{figure*}[ht]
\begin{center}
\resizebox{0.37\hsize}{!}{\includegraphics*{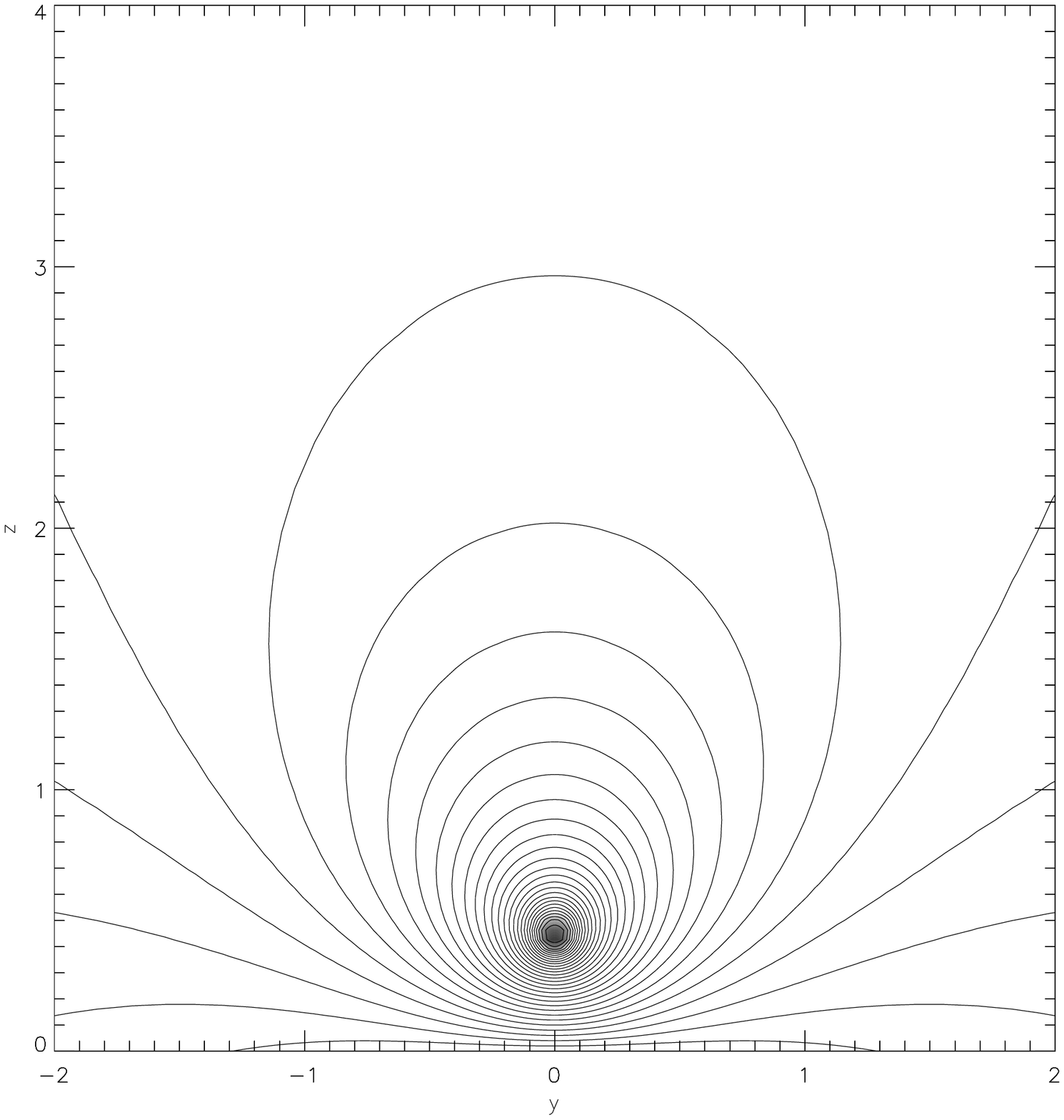}}
\end{center}
\caption{Magnetohydrostatic isothermal equilibrium with normal topology.}
\label{normalsole}
\end{figure*}

In the non-force-free models of this paper calculated in the Cartesian half-space $z>0$, there are two current systems: one centered at the center of the flux rope associated with the axial magnetic flux and one associated with the body of plasma centered in the bottom half of the flux rope.  From a distance, these current systems resemble line currents to lowest order, and their interaction determines the global field topology and far field structure of a solution. Thus the topology can be controlled via the signs and ratio of these currents.  In inverse-topology cases the shear- and plasma-induced currents are flowing in the same axial direction.  Normal-topology cases have additional complexity because the axially-directed currents associated with the shear and the plasma forces are flowing in opposite directions.

An example is shown in Figure~\ref{normalsole}.  FINESSE is less suited to the calculation of normal prominences than of inverse prominences because it is designed to find equilibria with $\psi =0$ at the magnetic axis and $\psi >0$ everywhere else, with maximum on the boundary.  The flux function of a normal prominence has a local minimum at the magnetic axis but, since the overlying bipolar arcade and the lower half of the flux rope must have flux traveling in the same horizontal direction, the flux function must have a saddle point above the flux rope, above which the flux function must decrease with height (see Figure~\ref{topologies}).  The example in Figure~\ref{normalsole} has such a saddle point far above the flux rope outside the field of view of the plot.  It is not yet clear how to incorporate this and other topological complications in a controlled way.

\section{Discussion}
\label{discussion}

A major obstacle in the way of understanding solar activity is the difficulty of capturing in models the complexity of the plasma dynamics and of the magnetic field structure.  For this reason, most recent efforts to model the solar atmosphere have attempted to recreate as far as practically possible the full 3D geometrical complexity of the physical parameters (e.g. Amari et al.~2003a, 2003b, Roussev et al.~2003, Wiegelmann \& Neukirch~2006).  Here we have adopted the alternative approach, following Low \& Hundhausen~(1995) and Low \& Zhang~(2004), of studying a simple generic physical system: a single solar prominence plasma enhancement suspended in a near-potential coronal magnetic field. 

We have given new numerical magnetohydrostatic solutions describing the gravitationally stratified, bulk equilibrium of cool, dense prominence plasma embedded in the near-potential coronal field.  These solutions are calculated using the FINESSE magnetohydrodynamics equilibrium solver and describe the morphologies of magnetic field distributions in and around prominences and the cool prominence plasma that these fields support.  The new solutions were not accessible by previous analytical techniques.  The numerical method allows us to prescribe in a controlled way the temperature or entropy as a function of the magnetic flux function, enabling flexibility of choice of the physical parameter distributions.  We focussed on new solutions with a range of values of the temperature, the magnetic shear, the polytropic index and with large temperature variations perpendicular to the magnetic field.  The axial component of the magnetic field gave demonstrably increased structural integrity to that field. With temperature a flux function and a low-density ambient atmosphere, a temperature minimum and a plasma pressure maximum at the flux rope center were found to produce a relatively evacuated cavity within which the cool plasma was embedded.  Such a cavity has been seen many times in observations.  For solutions with entropy a flux function, the polytropic index and the temperature distribution were found to be related in a simple way for separable cases.

The stability properties of these new equilibria can be determined in the form of a full resistive MHD spectrum by solving the linearized MHD equations using a companion hyperbolic stability solver, PHOENIX (Blokland et al.~2007a).  Because of the centrality of prominences and their fields to the most geoeffective space weather phenomenon, the CME, precise knowledge of these stability properties is very desirable.  Measurement of the spectrum of MHD waves, called MHD spectroscopy in analogy with quantum mechanical spectroscopy, is still in its infancy but may well lead to a firm knowledge of the internal characteristics of plasmas structures (Goedbloed 2002).  MHD spectroscopy entails a separate study of the nonlinear static equilibrium configuration on the one hand and the various linear wave structures that can occur on the other.  The first study is the subject of this paper and the spectroscopy will be treated in a sequel.  The new equilibria also serve as suitable starting points for time-dependent MHD simulations using, e.g., the Versatile Advection Code (VAC).

A 3D extension of this work is possible via the study of helically symmetric MHS configurations, which are also governed by a Grad-Shafranov equation with a linear elliptic operator.  Bearing in mind the difficulty of calculating 3D MHS equilibria of mixed elliptic-hyperbolic type it is tempting to settle for a simple option: that of abandoning the direct calculation of equilibria and their linear perturbation spectra and exclusively working on time-dependent MHD simulations instead.  This would be an inferior option since it passes up the precision and detail offered by the equilibrium and spectral theory.  Optimally one would hope to exploit both approaches in tandem.


\acknowledgements
We thank the referee for helpful and constructive comments.  This work was conducted while GP was a participant in the National Aeronautics and Space Administration (NASA) Postdoctoral Program at Goddard Space Flight Center, and was based at National Solar Observatory, Tucson.  GP thanks RK for kind hospitality during a research visit to Rijnhuizen.


\appendix

\section{Appendix}
\subsection{Numerical solution of the problem}

Suppose we seek only separable solutions to Equation~(\ref{reducedfb}).  In this case solutions of Equation~(\ref{hydrostatic}) for $p$ and $\rho$ must have identical $\psi$-dependence.  Let 

\begin{equation}
p(\psi ,z) = p_0 (\psi )\tau_p (z) ,
\end{equation}

\noindent where $\tau_p (z)$ describe the field-aligned variation of the pressure with height.  Then Equation~\ref{reducedfb} takes the form

\begin{equation}
\nabla^2\psi = -A[F(\psi )-B\tau_p (z)\Pi (\psi)]\label{coreeq},
\end{equation}

\noindent where

\begin{eqnarray}
AF(\psi ) & = & \frac{1}{2}\frac{df^2(\psi )}{d\psi},\\
AB\Pi (\psi ) & = & -4\pi \frac{dp_0 (\psi )}{d\psi},
\end{eqnarray}

\noindent and $A$ and $B$ are constants: $B$ is related to the plasma $\beta =8\pi p/|{\bf B}|$ and is specified at input, while $A$ is to be determined by the numerical algorithm as an eigenvalue of the problem.

The weak form of equation~(\ref{coreeq}) is

\begin{equation}
\int_V \chi\nabla^2\psi dV = -A\int_V \chi [F(\psi )-B\tau_p (z)\Pi (\psi)]dV
\end{equation}

\noindent
or

\begin{equation}
\int_V \nabla\chi\cdot\nabla\psi dV = A\int_V \chi [F(\psi )-B\tau_p (z)\Pi (\psi)]dV+\int_{\partial V} \chi\partial_n\psi dS .\label{weakform}
\end{equation}

\noindent
for arbitrary test functions $\chi$.  Since this is true for any $\chi$, we may transform $\chi\rightarrow\delta\chi$ in the equation above to obtain the variational form.

We represent $y$, $z$, $\psi$, $\chi$ by bicubic isoparametric elements defined in a rectangular coordinate system $(s,t)\in [-1,1]^2$.  Within a cell, any function $f(y,z)$ has finite element representation

\begin{eqnarray}
f(y(s,t),z(s,t)) & = & \sum_{s_0,t_0} \left[ H_{00}(s,t)f(y_0,z_0)+H_{10}(s,t)\frac{\partial f}{\partial y} (y_0,z_0)+H_{01}(s,t)\frac{\partial f}{\partial z} (y_0,z_0)\right.\nonumber\\
& & \left.+H_{11}(s,t)\frac{\partial^2 f}{\partial y\partial z} (y_0,z_0)\right] ,
\end{eqnarray}

\noindent
where $H_{00}, H_{10},\dots$ are the usual Hermite bicubic polynomials.  The summation runs over corner points $(s_0,t_0)$ of the cell and $y_0=x(s_0,t_0)$, $z_0=y(s_0,t_0)$.  Also, $\psi$, $y$ and $z$ are approximated by the same locally defined interpolating functions

\begin{equation}
\psi (s,t)=\sum_{s_0,t_0} \left[ H_{00}(s,t)\psi (s_0,t_0)+H_{10}(s,t)\frac{\partial\psi}{\partial s} (s_0,t_0)+H_{01}(s,t)\frac{\partial\psi}{\partial t} (s_0,t_0)+H_{11}(s,t)\frac{\partial^2\psi}{\partial s\partial t} (s_0,t_0)\right] ,
\end{equation}

\noindent
etc.  The Picard iteration of the variational form of equation~(\ref{weakform}) leads to an iterative matrix problem of the form $K_{ij}x_j^{(n+1)}=b_i^{(n)}$.  Here

\begin{equation}
K_{ij}=\int_0^1\int_0^1 \nabla H_i(s,t)\cdot\nabla H_j(s,t) J dsdt ,\label{cartkij}
\end{equation}

\noindent
the $x^{(n+1)}$ are the coefficients of the interpolating functions of the finite elements and

\begin{equation}
b_i^{(n)}=A\int_0^1\int_0^1 H_i [F(\psi^{(n)})-B\tau_p (z)\Pi(\psi^{(n)})]Jdsdt,\label{cartbi}
\end{equation}

\noindent
where $J=\frac{\partial (y,z)}{\partial (s,t)}$ and $H_i$ are the finite elements, given single subscripts for simplicity.

\subsection{Scaling}

We now introduce scale parameters which may be used to normalize the physical quantities.  If $B_0$ is a typical magnetic field strength, all quantities can be made dimensionless as follows:

\begin{eqnarray}
\bar{y} & = & y/H,\\
\bar{z} & = & z/H,\\
\bar{f} & = & \delta f/B_0,\\
\bar{p} & = & \delta^24\pi p/B_0^2,
\end{eqnarray}

\noindent
Here $\delta =\frac{HB_0}{\psi_0}$ is a dimensionless parameter that enters when we choose to normalize the magnetic flux $\bar{\psi} =\psi /\psi_0$ where $\psi_0$ is a typical $\psi$ value.  In practice we only work with dimensionless quantities and drop the bars without confusion.

\end{document}